\documentclass[aps,prb,twocolumn,showpacs,superscriptaddress,groupedaddress,floatfix]{revtex4-1}
\usepackage{color}
\usepackage{mathtools}
\usepackage{latexsym}
\usepackage{graphicx}
\usepackage{float}
\usepackage{dcolumn}
\usepackage{bm}
\usepackage{amssymb}
\usepackage{amsmath}
\usepackage{mathtools}
\usepackage[space]{grffile}
\usepackage{xcolor}

\newcommand{\ket}[1]{| #1 \rangle}

\newcommand{\be}{\begin{equation}}
\newcommand{\ee}{\end{equation}}
\newcommand{\bea}{\begin{eqnarray}}
\newcommand{\eea}{\end{eqnarray}}

\def\nn{\nonumber\\}
\def\fr#1{(\ref{#1})}

\begin{document}


\title{Almost strong \(0,\pi\) edge modes in clean, interacting 1D Floquet systems}

\author{Daniel J. Yates$^{1}$}
\author{Fabian H.L. Essler$^{2}$}
\author{Aditi Mitra$^{1}$}
\affiliation{$^{1}$Center for Quantum Phenomena, Department of Physics, New York University, 726 Broadway, New York, NY, 10003, USA\\
$^{2}$Rudolf Peierls Centre for Theoretical Physics, University of Oxford, Oxford OX1 3PU, UK}

\date{\today}

\begin{abstract}
Certain periodically driven quantum many-particle systems in one
dimension are known to exhibit edge modes that are related to
topological properties and lead to approximate degeneracies of the
Floquet spectrum. A similar situation occurs in spin chains, where
stable edge modes were shown to exist at all energies in certain
integrable spin chains. Moreover, these edge modes were found to be
remarkably stable to perturbations. Here we investigate the stability
of edge modes in interacting, periodically driven, clean systems. We
introduce a model that features edge modes that persist over times
scales well in excess of the time needed for the bulk of the system to
heat to infinite temperatures.
\end{abstract}
\maketitle

\section{Introduction}
Non-abelian edge modes have attracted considerable attention as a possible
route to quantum information
processing~\cite{Kitaev06,NayakRMP08,Fendley09,Alicea12,Beenaker13}. Such
edge modes occur in the ground state sector of various models, and
information encoded in them is protected by a finite gap to
excitations. In a series of recent works~\cite{Fendley16, Fendley17,Nayak17,Garrahan18,Garrahan19}
it was established that, remarkably, certain spin models support
topological edge modes at all energy densities that are either stable
or very long-lived. Stable edge modes were termed \emph{strong zero
modes} in Ref.~\onlinecite{Fendley16} and are a reflection of the existence of an
operator $\Psi_0$ that commutes with the Hamiltonian \(H\) in the
thermodynamic limit, anti-commutes with a discrete (say
\(\mathbb{Z}_2\)) symmetry of the Hamiltonian \(\mathcal{D}\), \(\{\Psi_0,\mathcal{D}\}=0\),
and is normalizable \(\Psi_0^2=O(1)\). The presence of a strong zero
mode implies a parameter regime where the entire spectrum of the
Hamiltonian is approximately doubly degenerate, with the almost
degenerate eigenstates being \(\{|n\rangle,
  \Psi_0|n\rangle\}\), and correspond to two different discrete
symmetry sectors.

Strong zero modes were shown to exist in the
transverse field Ising model, which has a free fermionic spectrum, and
in the XYZ spin chain~\cite{Fendley16}, which is an interacting integrable
theory. Importantly, these edge features were shown to be extremely
robust to perturbations about these limits in the sense that
\emph{almost strong zero modes} with long but finite life times
persist~\cite{Fendley17}. Edge modes that lead to approximate
degeneracies at all energies are also known to occur in periodically driven
systems~\cite{Sen13,Bahri15,Khemani16,Para16,Yao17,Potter18} and are
closely related to symmetry-protected topological (SPT)
phases~\cite{Wen17,Chamon15,Kitagawa10,Potter16,Else16,Roy16,Sondhi16a}.

By virtue of the periodicity of the spectrum of the (stroboscopic) time
evolution operator $U(T)$ the resulting structure of edge modes is richer
than in the equilibrium case: in addition to (almost) zero energy
modes there are so-called $\pi$-modes, which correspond to a
quasi-energy $\epsilon\approx\pm\pi/T$, where $T$ is the period of the
drive. In the terminology introduced above this corresponds to the
existence of two operators $\Psi_0$ and $\Psi_{\pi}$ that are
normalizable, \(\Psi_{0,\pi}^2 =O(1)\), anti-commute with a discrete
symmetry of the system \(\{\Psi_{0,\pi},\mathcal{D}\}=0\) and respectively approximately
commute \(\left[\Psi_0,U(T)\right]\approx 0\) or anticommute
\(\{\Psi_{\pi},U(T)\}\approx 0\) with the time evolution operator $U(T)$. In
terms of the spectrum of the Floquet Hamiltonian the existence of
$\Psi_0$ implies the presence of pairs of almost degenerate
eigenstates $\{|n\rangle,\Psi_0|n\rangle\}$, while the existence of
$\Psi_\pi$ implies the existence of pairs of eigenstates
$\{|n\rangle,\Psi_\pi|n\rangle\}$ whose energies (approximately)
differ by $\pi/T$.

The existence of strong zero and $\pi$ mode operators in non-interacting
periodically driven models~\cite{Sen13}, in the high-frequency
limit~\cite{Chamon15}, and in the Floquet-many body localization
context~\cite{Chandran14,Bahri15,Khemani16,Yao17,Potter18}, has been studied. In
the high-frequency regime the Floquet Hamiltonians typically studied in the
literature become short-ranged and the situation becomes very similar to the
equilibrium case~\cite{Fendley17}. The question of what happens in interacting,
clean Floquet systems away from the high-frequency regime has not yet been
explored in any detail. In Ref.~\onlinecite{Para16} it was shown that edge modes
lead to approximate degeneracies in the Floquet spectrum of a particular clean,
interacting system. However the implications of this for the dynamics of the
modes and their robustness to heating was not investigated.

Periodically driven clean systems are known
to heat up~\cite{Lazarides14,Hyungwon14,DAlessio14,Ponte15,Haldar18}
and are generically characterized by Floquet Hamiltonians with long-ranged
interactions, so that one would not expect long-lived edge modes to
exist at all energy densities. We show that in contrast to this
expectation there exist periodically driven interacting systems
that feature almost strong zero and $\pi$ modes at all energy
densities, even though the system heats on much shorter time scales.

The paper is organized as follows. Section~\ref{binary} presents results for
the strong modes for a free Floquet system. Section~\ref{ternary} presents
results for the almost strong modes of the interacting Floquet system.
Section~\ref{FloquetH} derives effective interacting Floquet Hamiltonians
around some exactly solvable limits, and compares almost strong modes
obtained from them to that obtained from the full time-evolution.
Section~\ref{conclu} presents the conclusions. The details of the analytic
calculations and additional discussions are relegated to the Appendices.

\section{Strong zero and $\pi$ modes for the free binary drive} \label{binary}
It is
instructive to explicitly construct the strong zero and $\pi$ mode
operators for periodically driven systems with Floquet Hamiltonians
that can be expressed as fermion bilinears. As an example we consider
an Ising binary drive which switches between two Hamiltonians for
equal durations \(T/2\)~\cite{Sen13, Khemani16, Sondhi16a,Gritsev17},
\bea
U(T)&=& e^{\frac{-iTJ_x}{2} H_{xx} }e^{\frac{-iT\mu}{2}H_{z}}\ ,\nn
H_{xx} &=& \sum_{i = 1}^L \sigma^x_i \sigma^x_{i+1} \ ,\quad
H_z=\sum_{i = 1}^L \sigma^z_i\ .
\label{binarydrive}
\eea
In the following we set \(J_x =1\). The model \fr{binarydrive} has a
$\mathbb{Z}_2$ symmetry of rotations around the z-axis by 180 degrees,
generated by \(\mathcal{D} = \sigma_1^z \sigma_2^z \dots
\sigma_L^z\). We now construct 
operators $\Psi_{0,\pi}$ that are localized at the boundaries such that
\(U^\dagger(T) \Psi_{0,\pi} U(T) = \pm \Psi_{0,\pi}\), with an error
that is exponentially suppressed in the system size \(L\). It is
convenient to introduce Majorana fermions
$a_{2\ell-1}=\prod_{j=1}^{\ell-1}\sigma^z_j\sigma^x_\ell$ and
$a_{2\ell}=\prod_{j=1}^{\ell-1}\sigma^z_j\sigma^y_\ell$, and collect
the even and odd labeled Majoranas into two vectors
\(\vec{a}_{\rm odd}=(a_1,a_3,\ldots,a_{2L-1}), \vec{a}_{\rm
  even}=(a_2,a_4,\ldots,a_{2L})\). Both $H_{xx}$ and $H_z$ are
quadratic in the Majorana operators, and concomitantly their
stroboscopic time evolution can be cast in the form
\be
\begin{pmatrix}
\vec{a}_{\rm odd}\big((n+1)T\big)\\ \vec{a}_{\rm even}\big((n+1)T\big)
\end{pmatrix}= M\begin{pmatrix}
\vec{a}_{\rm odd}\big(nT\big)\\ \vec{a}_{\rm even}\big(nT\big)
\end{pmatrix},
\ee
where $M$ is an orthogonal matrix, and is given in Appendix~\ref{appA}. We then
make the Ansatz $\Psi_{0,\pi}=\sum_j \varphi^{(0,\pi)}_j\ a_j$ for the
zero/$\pi$-mode operators and require them to be invariant (up to a sign
in case of the $\pi$-mode) under stroboscopic time evolution. This
leads to an eigenvalue equation of the form
$\varphi_j^{(\sigma)}=\cos(\sigma)\sum_\ell
M_{j\ell}\varphi_\ell^{(\sigma)}$.
Interestingly, these equations can essentially be solved in closed
form (Appendix~\ref{appA}) in the limit of large system size $L$.

\begin{figure}[ht]
\includegraphics[width = 0.45\textwidth]{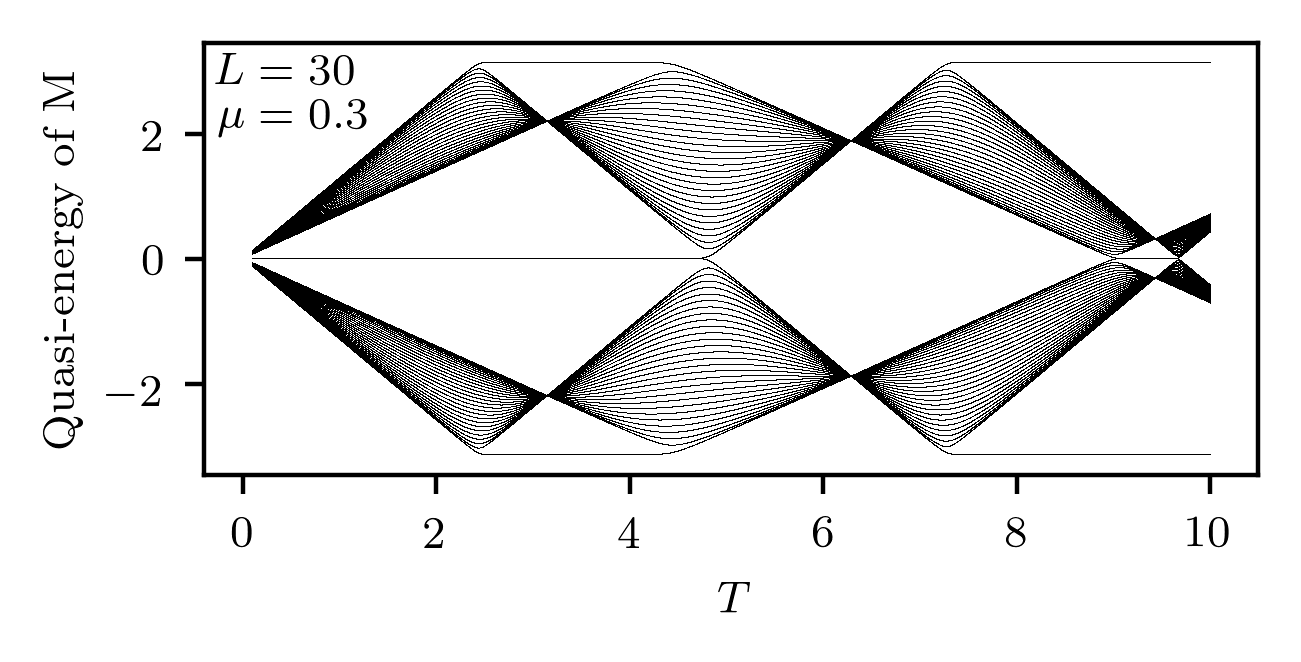}
\caption{\label{fig1}
Quasi-energies obtained from the eigenvalues of $M$ for the free binary
drive. Different topological phases are visible as the drive period \(T\)
is varied. From left to right, the phases are $M_0$, $M_0+M_\pi$, trivial,
and $M_\pi$, where $M_\sigma$ indicates the existence of a strong
Majorana edge mode, \emph{cf.} Eqn.~\eqref{szm}.
}
\end{figure}

Dropping
contributions that are exponentially small in system size, the
operators can be written in the form
$\Psi_\sigma\approx\Psi^L_{\sigma}+\Psi^R_{\sigma}$, where
$\Psi_{\sigma}^L$ ($\Psi^R_{\sigma}$) has support mainly near the
left (right) boundary, where
\bea
\Psi_{0}^L &=&  \sum_{j \geq 1}
\epsilon_-^{j-1}
\left[\cos\Big(\frac{T\mu}{2}\Big) a_{2j-1}
-\sin\Big(\frac{T\mu}{2}\Big) a_{2j}\right],\nn
\Psi_{\pi}^L &=& \sum_{j \geq 1} \epsilon_+^{j-1} \left[
  \sin\Big(\frac{T\mu}{2}\Big) a_{2j-1} + \cos\Big(\frac{T\mu}{2}\Big)
  a_{2j}\right].\label{szm} \eea Here we have defined
$\epsilon_-=\tan(\frac{T\mu}{2})\cot(\frac{TJ_x}{2})$ and
$\epsilon_+=-\cot(\frac{T\mu}{2})\cot(\frac{TJ_x}{2})$.
Similar \(0,\pi\) mode operators appear in Ref.~\onlinecite{Sen13}
for a time-symmetrized version of \(U(T)\).
Both modes can be readily seen to
anticommute with the generator of the $\mathbb{Z}_2$ symmetry
\(\{\mathcal{D},\Psi_{0,\pi}\} = 0\), which establishes that acting with
$\Psi_{0,\pi}$ on an eigenstate of $U(T)$ that is even (odd) under the
$\mathbb{Z}_2$ symmetry gives an eigenstate $U(T)$ that is odd (even). The
condition for $\Psi_\sigma$ to be normalizable in the thermodynamic limit is
$|\epsilon_\sigma|<1$ and $|\epsilon_\sigma|=1$ fixes the location of the
topological phase transitions of the model \emph{cf.} Fig.~\ref{fig1}. Here the
topological phases are that of a free BDI Floquet
SPT~\cite{Kitagawa10,Potter16,Else16,Roy16,Sondhi16a} with an invariant in
\(\mathbb{Z}\times \mathbb{Z}\), the two integers being the numbers of \(0,\pi\)
edge modes. The drive used in this paper only generates indices of \(0\) or
\(1\) for each edge mode species so that the difference between \((\mathbb{Z}_2
\times \mathbb{Z}_2)\) and \((\mathbb{Z} \times \mathbb{Z})\) is not apparent.
More general drives that preserve the BDI symmetries can realize a larger
numbers of edge modes in both species \cite{Sen13,Delplace14,Yates17,Yates18}.
Whether these additional edge modes are associated with additional strong mode
operators is left for future study.

In the \(T \rightarrow 0\) limit, we can perform a high-frequency
expansion~\cite{Polkov13,Eckardt15}
to leading order and obtain the Floquet Hamiltonian
\(H_F=\frac{1}{2}\left( J_x  H_{xx} + \mu H_{z} \right)\), which is a
transverse field Ising model. In this limit our expression \eqref{szm}
for the zero mode $\Psi_0$ reduces to that previously
obtained in equilibrium~\cite{Kitaev06,Fendley16,Nayak17}.
It is instructive to consider the strong edge modes in some simple
limiting cases~\cite{Sondhi16a}.
\begin{enumerate}
\item[1.]{\(T \mu = (2n + 1)\pi\) and \(T J_x\) arbitrary}

Here \(\exp(-i T\mu H_z/2)=(-i(-1)^n)^L\mathcal{D}\) and \(\sigma_1^x\)
becomes a strong \(\pi\) mode, while there is no strong zero
mode. This is consistent with \(\epsilon_-\rightarrow\infty\), which
signals to non-normalizability of our zero mode solution.
Only the first term in the expansion of \(\Psi_\pi^L\) in
Eq.~\eqref{szm} is non-zero and gives $\psi_\pi^L=\sigma_1^x$.
\item[2.]{\(T \mu = 2n \pi\) and \(T J_x\) arbitrary}

In this case we have \(\exp(-i T\mu H_z/2) = 1 \) and \(\sigma_1^x\)
becomes a strong zero mode, whereas there is no strong \(\pi\)
mode. This corresponds to the limit \(\epsilon_{+}\to\infty,
\epsilon_-\to 0\) in \fr{szm}.
\item[3.]{\(T J_x = (2n+1) \pi\) and \(T \mu\) arbitrary}

Then \(\exp(-i T J_x H_{xx}/2)=(-i (-1)^n)^{L-1}\sigma_1^x \sigma_L^x\) and it is
straightforward to check that both a strong 0 and \(\pi\) mode
exist. Their explicit expressions are given by the $j=1$ terms in \fr{szm}.
\item[4.]{\(T J_x = 2n \pi\) and \(T \mu\) arbitrary}

Here we have \(\exp(-i T J_x H_{xx}/2)= 1\) and no strong edge modes
exist unless \(T\mu/\pi\) is an integer. If this integer is odd (even) then
\(\sigma_1^x\) is a strong \(\pi\) (zero) mode.
\end{enumerate}

\section{Interacting ternary drive} \label{ternary}
\begin{figure}[ht]
\begin{minipage}{.99\columnwidth}
\includegraphics[width = \textwidth]{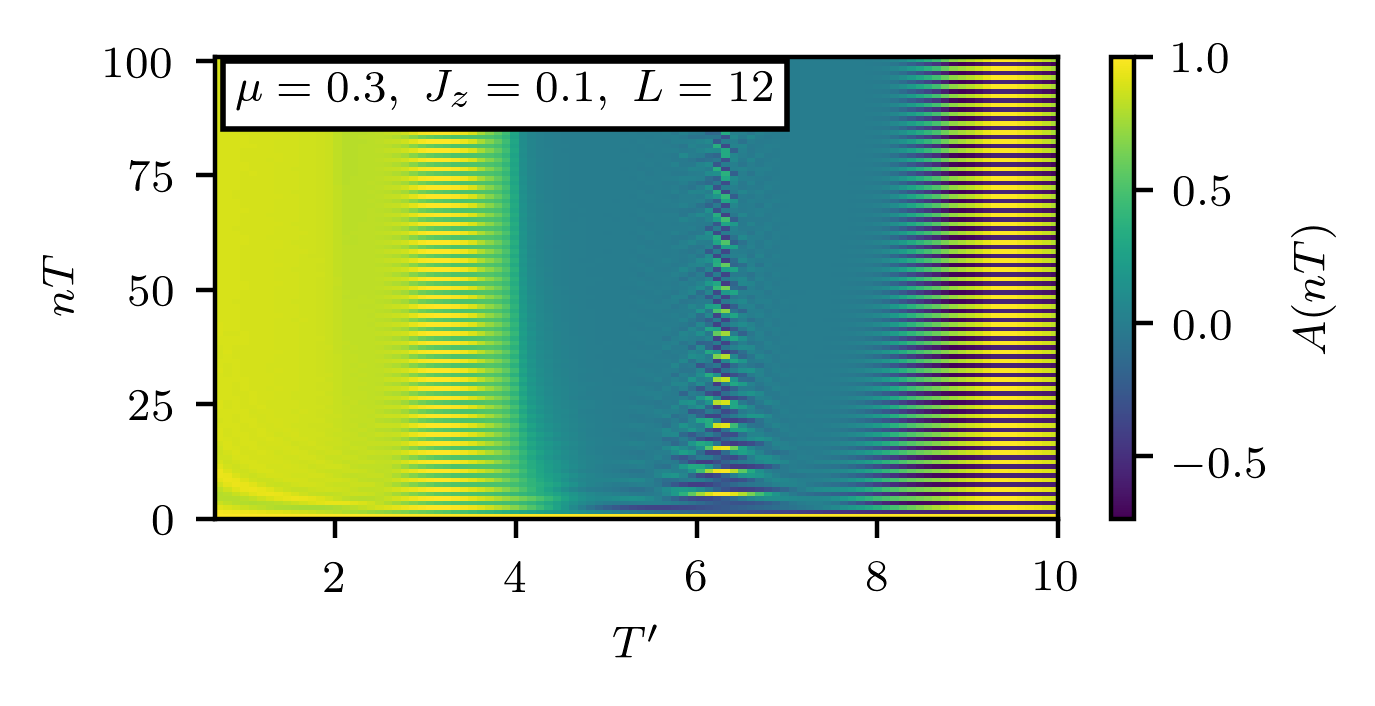}
\end{minipage}
\caption{\label{fig2} Edge mode diagnostic \(A_{}(nT)\) for the interacting
ternary drive with period \(T\) where \(T'=2T/3\). There are three parameter
regimes in which almost strong edge modes occur: $M_0$ (\(T'J_x<2\)),
$M_0+M_\pi$ (\(2<T'J_x<4\)) and $M_\pi$ (\(8<T'J_x\)). Here $M_{0,\pi}$
indicates the presence of a edge zero/$\pi$ mode.  The
structure seen at \(T' \approx 2\pi\) arises due to the flat band
section visible in Fig.~\ref{fig1} and the small value of \(J_z =
0.1\). A larger \(J_z\) would quickly dampen these oscillations. The
oscillatory behavior in the regime \(n\alt  10\) and \(T'\alt 2\) is a finite-size
effect: the system size, \(L = 12\), is too small for the spectrum to
``wrap around'' the unit circle for these high frequencies, \emph{cf.}
Fig.~\ref{fig11b}.  } 
\end{figure}

\begin{figure}[ht]
 \begin{minipage}{.99\columnwidth}
    \includegraphics[width = \textwidth]{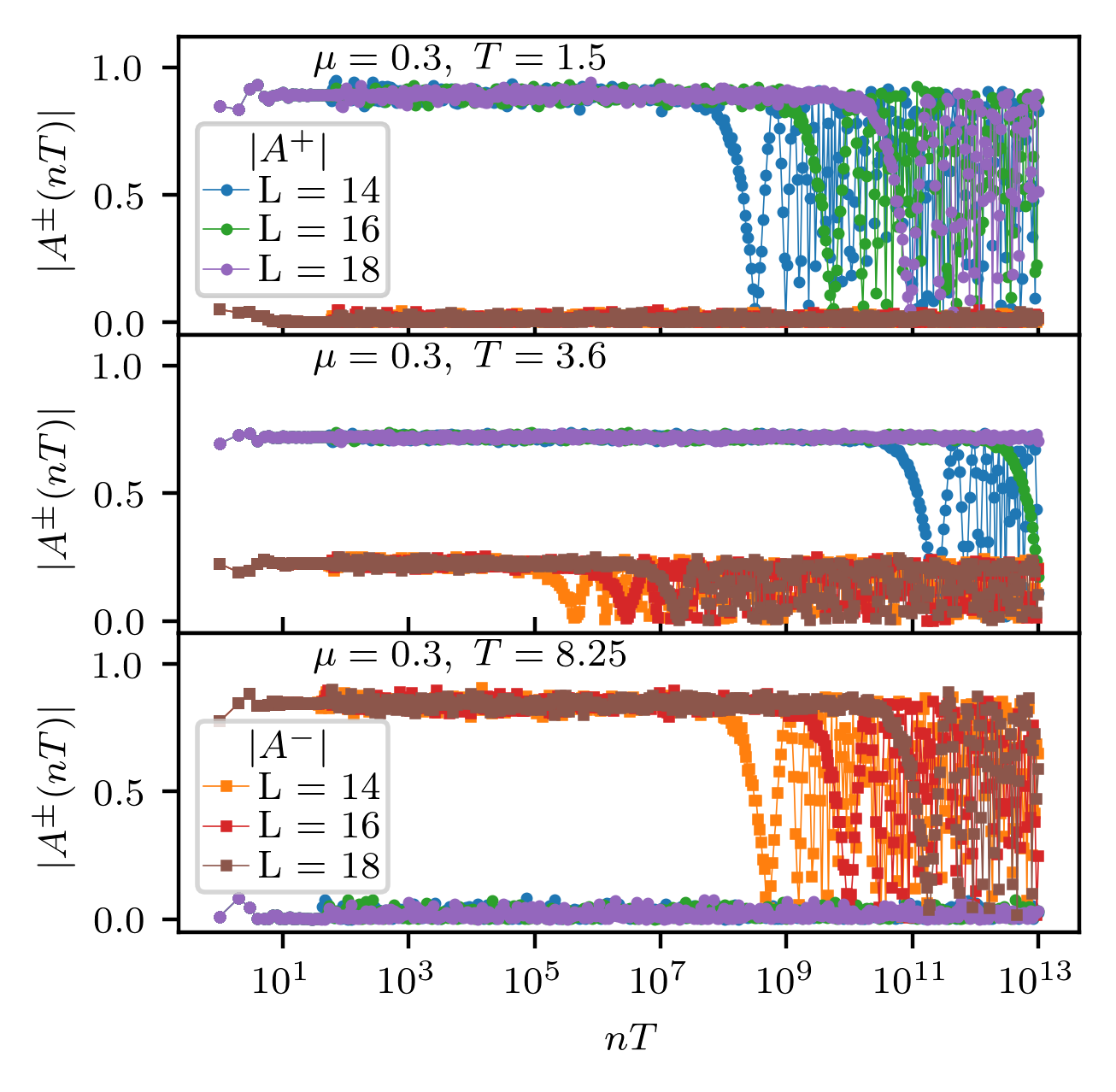}
    \end{minipage}
    \caption{\label{fig1b} Suitably symmetrized/anti-symmetrized overlaps
      \(A^{\pm}\) as a function of stroboscopic time \(nT\), for the
      binary drive. Top panel: a strong zero mode exists.
      Middle panel: a strong zero mode coexists with a strong \(\pi\) mode.
      Lower panel: Only a strong \(\pi\) mode exists. The lifetimes of
      the modes grow exponentially with system size. From top to
      bottom panels, all parameters are fixed, and only the drive
      frequency \(T^{-1}\) is decreased.}
\end{figure}

\begin{figure}[ht]
 \begin{minipage}{.99\columnwidth}
    \includegraphics[width = \textwidth]{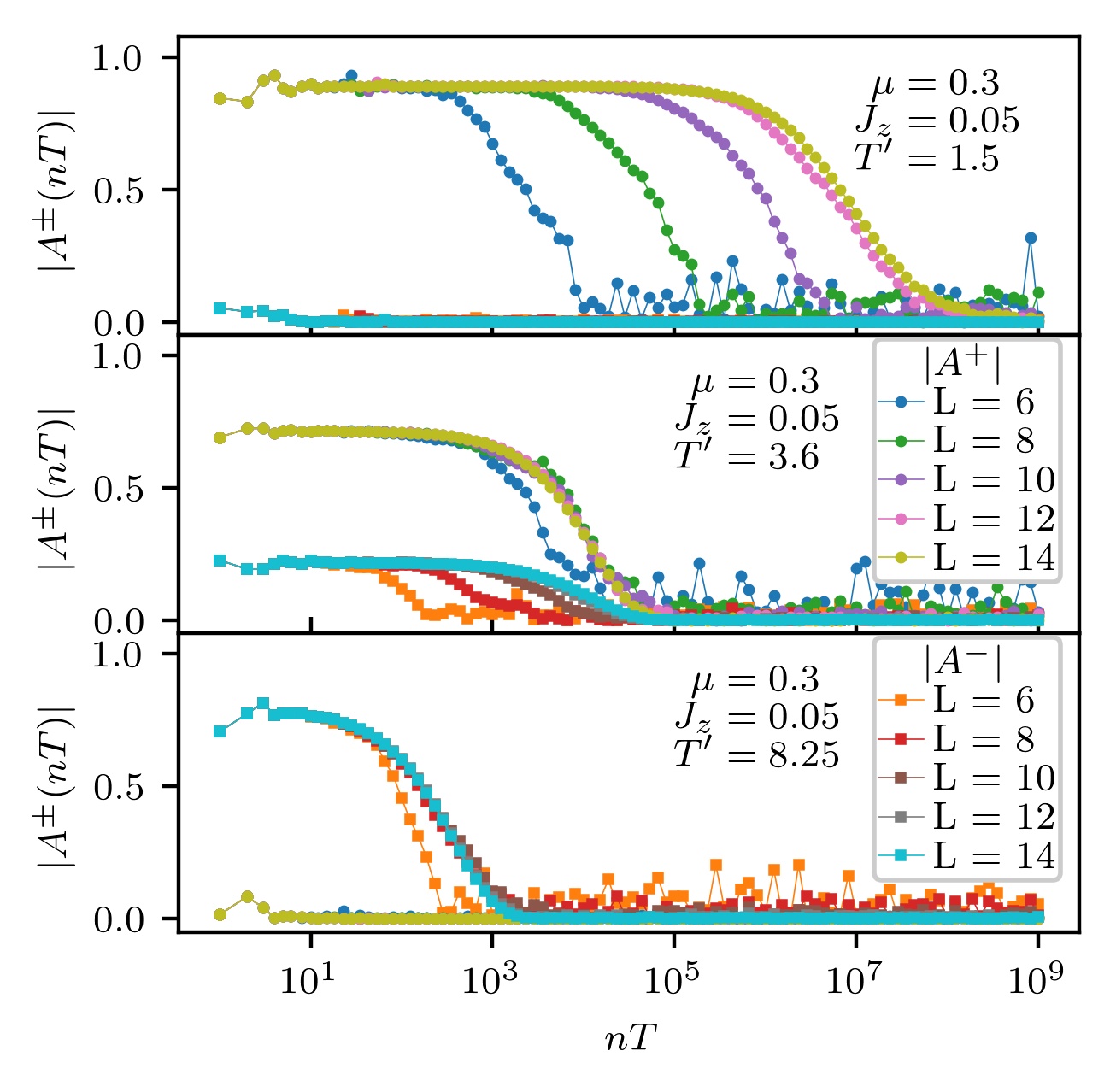}
    \end{minipage}
    \caption{\label{fig1c} Suitably symmetrized/anti-symmetrized overlaps
      \(A^{\pm}\) as a function of stroboscopic time \(nT\), for the interacting
      ternary drive with \(J_z/J_x=0.05\), and the same \(T',\mu\) as for the
      binary drive shown in Fig.~\ref{fig1b}. Note that the
        \(T'\) here equals the \(T\) in figure 3. Top panel: an almost strong
      zero mode exists whose lifetime grows with system size, and does not
      saturate for the sizes shown. Middle panel: almost strong zero and $\pi$
      modes coexist. The life time of the zero ($\pi$) mode saturates for system
      \(L=8\) (\(L=12\)). Lower panel: there is an almost strong \( \pi \) mode,
      whose lifetime saturates for system size \( L=8 \).}
\end{figure}

We now add interactions to the Floquet driving by dividing the period
into 3 equal parts
\begin{equation}
U(T) = e^{-i \frac{TJ_z}{3} H_{zz}} e^{-i \frac{TJ_x}{3}H_{xx}} e^{-i \frac{T\mu}{3}H_z},
\end{equation}
where \( H_{zz} = \sum_{i=1}^{L-1} \sigma_i^z \sigma_{i+1}^z\). We note that
\(\mathcal{D}\) remains a symmetry of this drive. We have studied this model by
means of exact diagonalization on system sizes up to \(L=14\). In the following
we set \(J_x=1\). It is useful to define \(T'\) as \(T'/2 = T/3\) so that when
\(J_z \rightarrow 0\) the ternary drive reduces to the solvable binary drive.
This facilitates comparisons between results for free (Figs.~\ref{fig1} and
\ref{fig1b}) and interacting drives (Figs.~\ref{fig2} and \ref{fig1c}). Guided
by the findings of Ref.~\onlinecite{Fendley17} in equilibrium we wish to
investigate the possible existence of almost strong zero and $\pi$ modes, i.e.
long-lived edge modes. In order to search for these modes we consider the
overlap of the boundary spin \(\sigma_1^x=a_1\) at time $nT$ with the boundary
spin at time zero
\begin{align}
A_{}(nT) = \frac{1}{2^L}{\rm Tr}\left[\sigma_1^x(nT)\sigma_1^x\right]= \frac{1}{2^L}{\rm Tr}\left[a_1(nT)a_1\right].
\end{align}
In the absence of any edge modes \(A_{}(nT)\) is expected to rapidly decay to
zero. On the other hand, almost strong zero or $\pi$ modes will have a
non-zero overlap with the edge spin \({\rm
  Tr}\left[\sigma^x_1\Psi_{0,\pi}\right]\neq 0\) and this prevents
\(A_{}(nT)\) from decaying to zero rapidly with time.  
The rationale behind these expectations is discussed in Appendix~\ref{appB}.

An alternative diagnostic of
edge modes is the autocorrelation function measured with respect to a
certain initial state \(|\psi\rangle\), defined as \(A_{\psi}(nT)
= \langle\psi|\sigma^x_1(nT)\sigma^x_1|\psi\rangle\). The
physical meaning of this quantity is that we start from an initial
state \(|\psi\rangle\), flip a spin at site 1, then time-evolve
until time \(nT\), and flip the spin back again obtaining a state
\(\sigma^x_1 U(nT) \sigma^x_1|\psi\rangle\). \(A_{\psi}(nT)\) then
measures the overlap of this state with one where the initial state
was evolved up to time \(nT\) without any spin-flips \(U(nT)
|\psi\rangle\). Thus this quantity measures the decoherence of any
edge mode. If almost strong modes exist, then after an initial
transient the two quantities \(A(nT),A_{\psi}(nT)\) behave
similarly (Appendix~\ref{appB}).

In Fig.~\ref{fig2} we show results for \(A(nT)\) as a function
of stroboscopic time $nT$ and drive period $T$ for parameters
$\mu=0.3$ and $J_z=0.1$. We see that edge modes persist for
considerable time even in the presence of interactions. For the
parameters shown, these modes are adiabatically connected to the free
case. In the remainder of the paper we analyze this behavior as a function
of system size \(L\), drive frequency \(T^{-1}\), and strength of
interactions \(J_z\).

Since the \(\pi\)-modes alternate sign every period, their persistence with time and
system size is most apparent in a staggered average over adjacent
stroboscopic times, \(A_{}^-(nT)=
\left[A_{}(nT)-A_{}((n+1)T)\right]/2\). It is similarly
convenient to extract the effects of zero modes by considering the
flat average \(A_{}^+(nT)=
\left[A_{}(nT)+A_{}((n+1)T)\right]/2\).
To set the stage we first investigate the behavior of \(A^{\pm}\) for
the free binary drive, where we know when strong edge modes exist.
In
Fig.~\ref{fig1b} we show results for $|A^\pm(nT)|$ for parameters
where (i) a strong zero
mode exists (top panel); (ii) strong zero and $\pi$ modes coexist
(middle panel); and
(iii) only a strong \(\pi\) mode exists. It is apparent from the top and
bottom panels that in the absence of the respective strong edge mode, the
corresponding diagnostic rapidly decays to zero, and this
behavior is system size independent. In contrast, when a strong edge mode
exists, the diagnostic stays constant on a time scale that grows with
system size. Fig.~\ref{fig1b} also reveals how the system rebounds
after the ``decay'', revealing recurrences characteristic of a free
system. The log scale of the \(x\)-axis masks the fact that the decays in
the free system are simple cosine oscillations that are exponentially
slow in system size.

We now turn to the ternary drive. Results for the edge mode
diagnostics for \(J_z/J_x=0.05\) are shown in Fig.~\ref{fig1c}. We observe
almost strong edge modes with life times that initially grow with
system size and eventually saturate.
We note that going from the top to the bottom panels the drive
frequency is being lowered, and this changes the life times of the
almost strong edge modes. In particular we see that for sufficiently
low frequency driving (middle and lower panels) the life times of the
almost strong modes saturate at increasingly lower system sizes.

An immediate question raised by the existence of long-lived edge modes
is whether they are related to some kind of prethermal
behavior~\cite{Abanin15,Bertini15,AMreview,Kuwahara2016,Bukov16,Mori16,Else17,Abanin17,Abanin17b}.
To answer this question we have investigated on what time scales
heating occurs in our system. We now show that the lifetime of the
modes far exceeds thermalization times by several orders of magnitude.

\begin{figure}[ht]
\includegraphics[width = .49\textwidth]{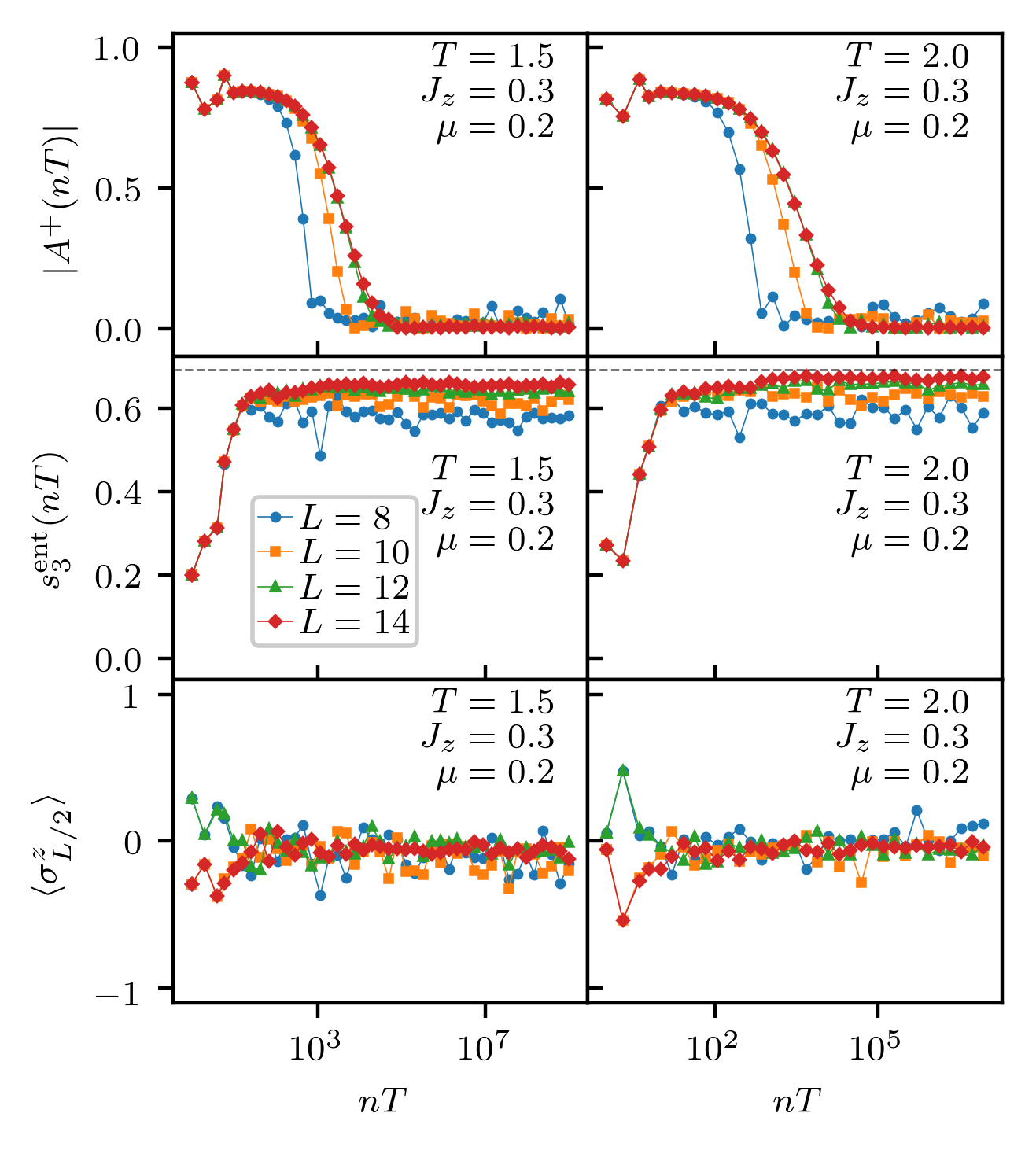}
\caption{\label{fig3}
Top panels: Time-evolution of edge zero mode diagnostic
\(A^+(nT)\) (\(A^-(nT)\approx 0, \,\text{not shown}\)).
Middle and lower panels: Time-evolution of the three-site
entanglement-entropy density and expectation value of the central spin
\(\langle \sigma^z_{j=L/2}(nT)\rangle\) for a system initialized in a classical
N\'eel state.
Middle and lower panels show thermalization on
time-scales that do not depend on system size, and occur on
time-scales much shorter than the lifetime of the almost strong mode
in the upper panel. }
\end{figure}

\begin{figure}
\includegraphics{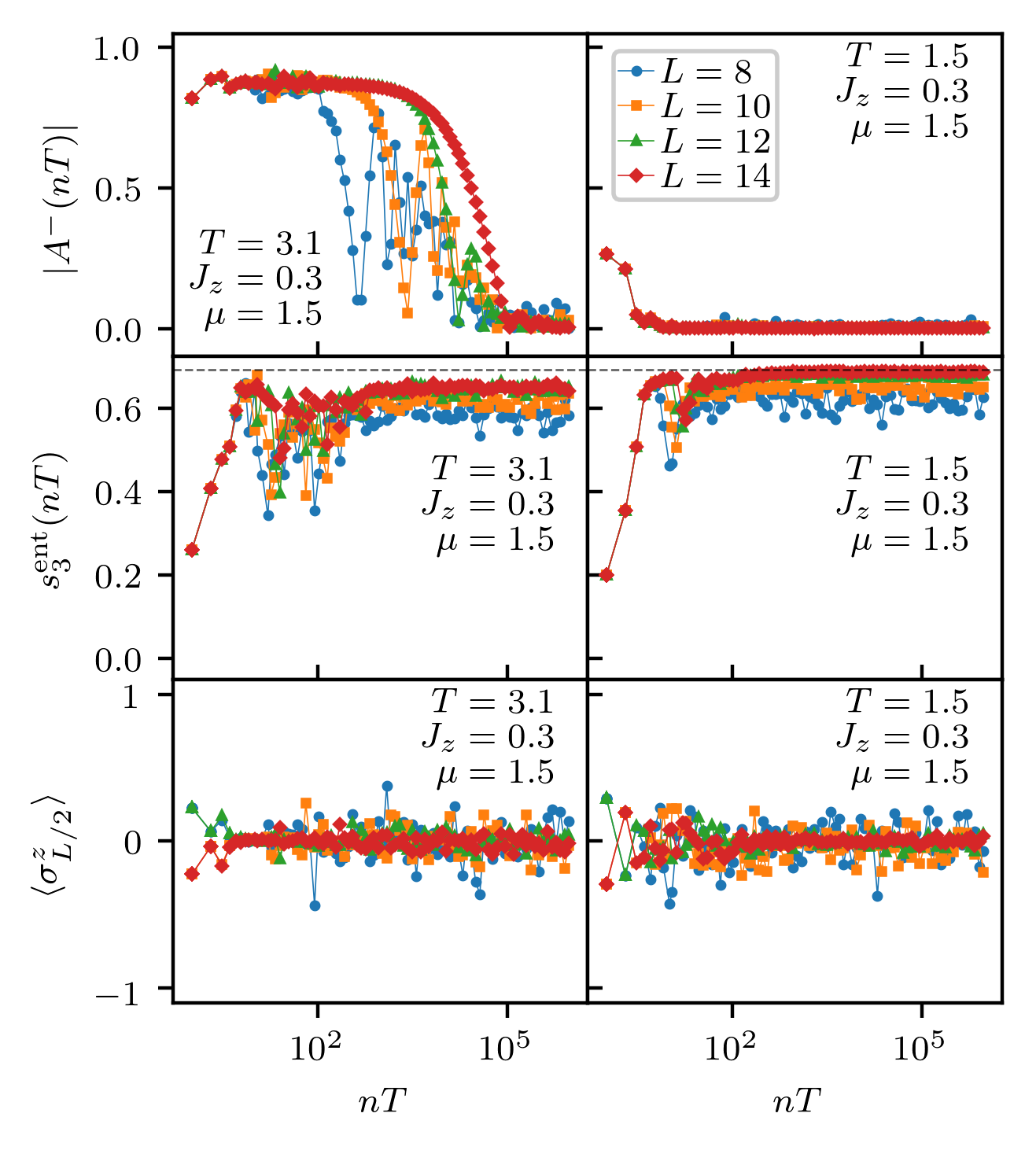}
\caption{\label{fig4} Top panels: almost strong $\pi$-mode diagnostic
\(A^-(nT)\) for two low frequency drives as a function of time. For
$T=3.1$ there is a long-lived $\pi$-mode up to times $nT\sim 10^5$,
while for $T=1.5$ there isn't. (\(A^+\approx 0, \,\text{for both cases, not shown}\))
Middle and bottom panels: time
evolution of the three-site entanglement-entropy density $s_3^{\rm
  ent}(nT)$ and average spin \(\langle
\sigma^z_{j=L/2}(nT)\rangle\) starting from a N\'eel state. The
system is seen to approach an infinite temperature state for
times of the order of $nT\sim 10$ for both parameter sets.
}\end{figure}

The comparison between the lifetime of almost strong modes, and thermalization
times are presented in Figs.~\ref{fig3} and~\ref{fig4}.
  Fig.~\ref{fig3} presents results for two different parameter points
  coinciding with the existence of almost strong zero modes,
  \(T = 1.5\) on the left panels, \(T = 2.0\) on the right panels, and \(\mu = 0.2, J_z =
  0.3\) for both.
In the top panels of
Fig.~\ref{fig3} we show the behavior of $|A^+(nT)|$ as a function of
stroboscopic time for several system sizes and parameters that correspond to two
different periods.
For these parameters \(A^-\approx 0\) within a cycle. We observe that
$|A^+(nT)|$ remains large for a substantial but finite time, indicating the
existence of an almost strong zero mode. For the parameters chosen in
Fig.~\ref{fig3}, system size of \(L=14\) is sufficient to show the saturation of
the lifetime with system size.

The lower two panels in Fig.~\ref{fig3} show the time-evolution of two measures
of thermalization, namely the entanglement entropy density for a subsystem of
size three and the expectation value of \(\sigma^z_{j=L/2}\) at the center of
the chain, both following a quantum quench from a N\'eel initial state. These
results show that the system heats to infinite temperature on a much shorter
time scale than the lifetimes of the edge modes. Note that at sufficiently late
times the entanglement entropy density approaches the infinite temperature limit
of \(\ln{(2)}\) (dashed line) as the system size is increased,
\emph{cf.} Fig.~\ref{finite1}. We focus on subsystem size
three as this is the maximal value for which finite-size effects (due
to the limited system size $L$) are sufficiently small.
We find that the behavior of the \(\sigma^z_j\) at
other positions is qualitatively similar in that it rapidly decays to zero,
including at the edge. We have considered several other initial states and
observed the same behavior.

\begin{figure}[ht]
\includegraphics[width = .49\textwidth]{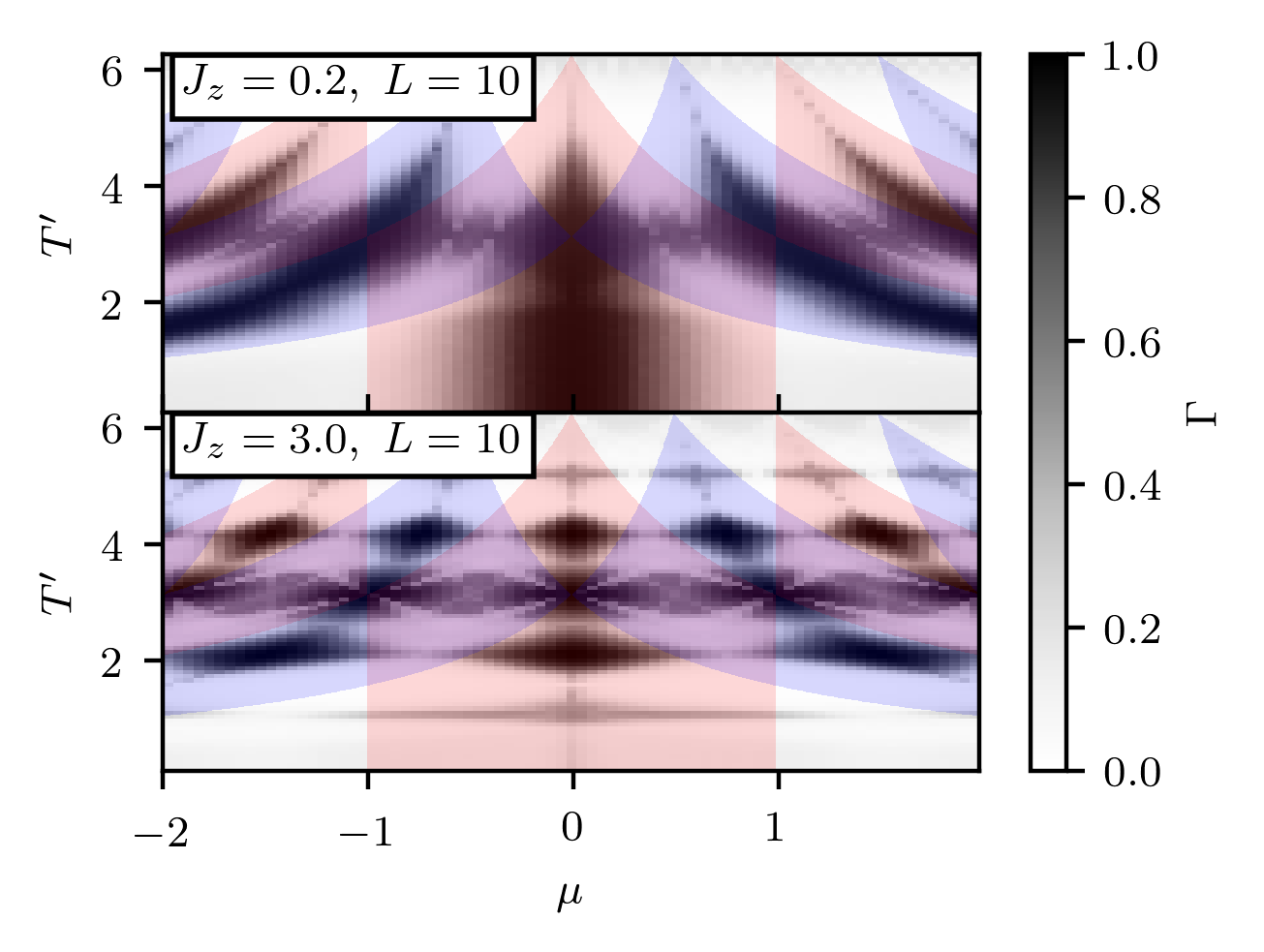}
\caption{\label{fig5}
Edge mode diagnostic $\Gamma$ for fixed \(J_z/J_x=0.2,3.0\). The black
regions indicate the presence of almost strong edge modes. For
comparison we also indicate where zero (red) and $\pi$-modes (blue)
exist in the binary drive model. In the top panel, the strong zero mode for \(|\mu|\le 1\)
and sufficiently small \(T\) is closely connected to the topological phase of the static Kitaev chain.
The blue and red wings for \(T>0\) are introduced by the Floquet driving and do not have a static analog. }
\end{figure}

In Fig.~\ref{fig4} we present analogous results for a parameter regime
in which an almost strong $\pi$ mode exists (left hand panels) and a
case in which there are no long-lived edge modes (right hand panels).
The results for the entanglement entropy density and the central spin
show that in both cases the system quickly heats to an infinite
temperature state. For $T=3.1$, $J_z=0.3$ and $\mu=1.5$ (left hand panel) the results for
$A^-(nT)$ reveal the existence of a $\pi$ edge mode long after
the system has thermalized. On the other hand, for $T=1.5$, $J_z=0.3$
and $\mu=1.5$ (right hand panel) the edge coherence disappears around the same time when
the system reaches an infinite temperature state.
We observe that upon decreasing \(J_z\) the lifetimes of
existing zero or $\pi$ modes will increase roughly as \(\sim
\exp(1/J_z)\). It is difficult to quantify this behavior more
precisely due to the limitations set by the system sizes accessible to us.
We typically find only a narrow parameter range in which \(J_z\) can
be varied while the lifetimes of zero/$\pi$ modes still saturate for
$L=14$. We find that the lifetimes of both zero and $\pi$ modes can be
extended  by moving closer to their respective integrable lines,
i.e. the centers of the blue and red regions In Fig.~\ref{fig5}.

A second diagnostic for detecting the presence of edge modes is the overlap of
\(\sigma^x_1\) between opposite symmetry sectors~\cite{Fendley17}. This is
defined as 
\be 
\Gamma=\frac{1}{2^L} \sum_s \max_{s'}|\langle s | \sigma_1^x |
s'\rangle|^2\ , 
\ee 
where $|s\rangle$ and $|s'\rangle$ denote the exact
eigenstates of the Floquet unitary $U(T)$. This diagnostic, since it takes a
mean of the overlap between opposite symmetry sectors, treats zero and $\pi$
modes on an equal footing. 
The reasoning why $\Gamma$ is
a useful edge mode diagnostic goes as follows. Up to corrections that
are exponentially small in system size strong edge modes
$\Psi_{0,\pi}$ map each eigenstate \(\ket{s}\) to another eigenstate
of opposite fermion parity, i.e. \(\ket{s'}\approx\Psi_{0,\pi}\ket{s}\).
As ${\rm Tr}(\sigma^x_1\Psi_{0,\pi})={\cal O}(1)$ strong edge modes
therefore lead to finite values of $\Gamma$. Reversing the argument,
the exponentially small factor $2^{-L}$ in the definition of $\Gamma$
can be compensated only if most eigenstates $|s\rangle$ of $U(T)$ have
a partner $|s'\rangle$ in the opposite symmetry sector such that
$|\langle s | \sigma_1^x | s'\rangle|^2={\cal O}(1)$. As almost all
eigenstates of $U(T)$ have finite correlation lengths this implies the
existence of fermionic edge modes.
We plot $\Gamma$ as a function of the parameters
$T'$ and $\mu$ of the ternary drive for fixed substantial interaction strengths
\(J_z/J_x=0.2,3.0\) in Fig.~\ref{fig5}. We observe almost strong edge modes
despite the Floquet Hamiltonian having sizeable interactions. For comparison we
show the regions in which strong edge modes exist in the binary drive. For the
system sizes accessible to us, the size and shape of the black regions that
indicate the presence of edge modes are only weakly affected by finite-size
effects (Appendix~\ref{appC}).

\subsection{Finite size effects} \label{finiteSec}
In a large finite system local thermalization \cite{Essler16} implies that
the difference between the time average of the system's reduced
density matrix $\overline{\rho_A}$, and an appropriate thermal reduced
density matrix $\rho_A^{\rm th}$, goes to zero as system size is
increased for a fixed choice of subsystem $A$
\be
|\overline{\rho_A}-\rho^{\rm th}_A|<\epsilon_L\ ,\quad \lim_{L\to\infty}\epsilon_L=0.
\ee
This means in particular that the time-averages of expectation values
of local observables, sufficiently far away from any boundaries,
approach thermal values as the system size is increased. We will now
show that our system quickly reaches an infinite temperature steady
state in this sense. Our discussion necessarily focusses on small
subsystems, and we are in particular unable to address questions such
as how the time scale at which the reduced density matrix of a large
subsystem (but still small compared to the system size $L$), approaches
its infinite temperature value within a given error, depends on the
size of the subsystem.
However, given that the correlation lengths in
our system are very short, all ``large'' observables are already
accessible in short subsystems. Considering how close a two point
function at separation $100$ is to its infinite temperature value is
essentially a purely academic question.

In the following we focus on
two representative local quantities, namely the \(z\)-component of the
spin in the centre of our chain and the entanglement entropies of
small subsystems. 
\begin{figure}[ht]
  \includegraphics[width=.49\textwidth]{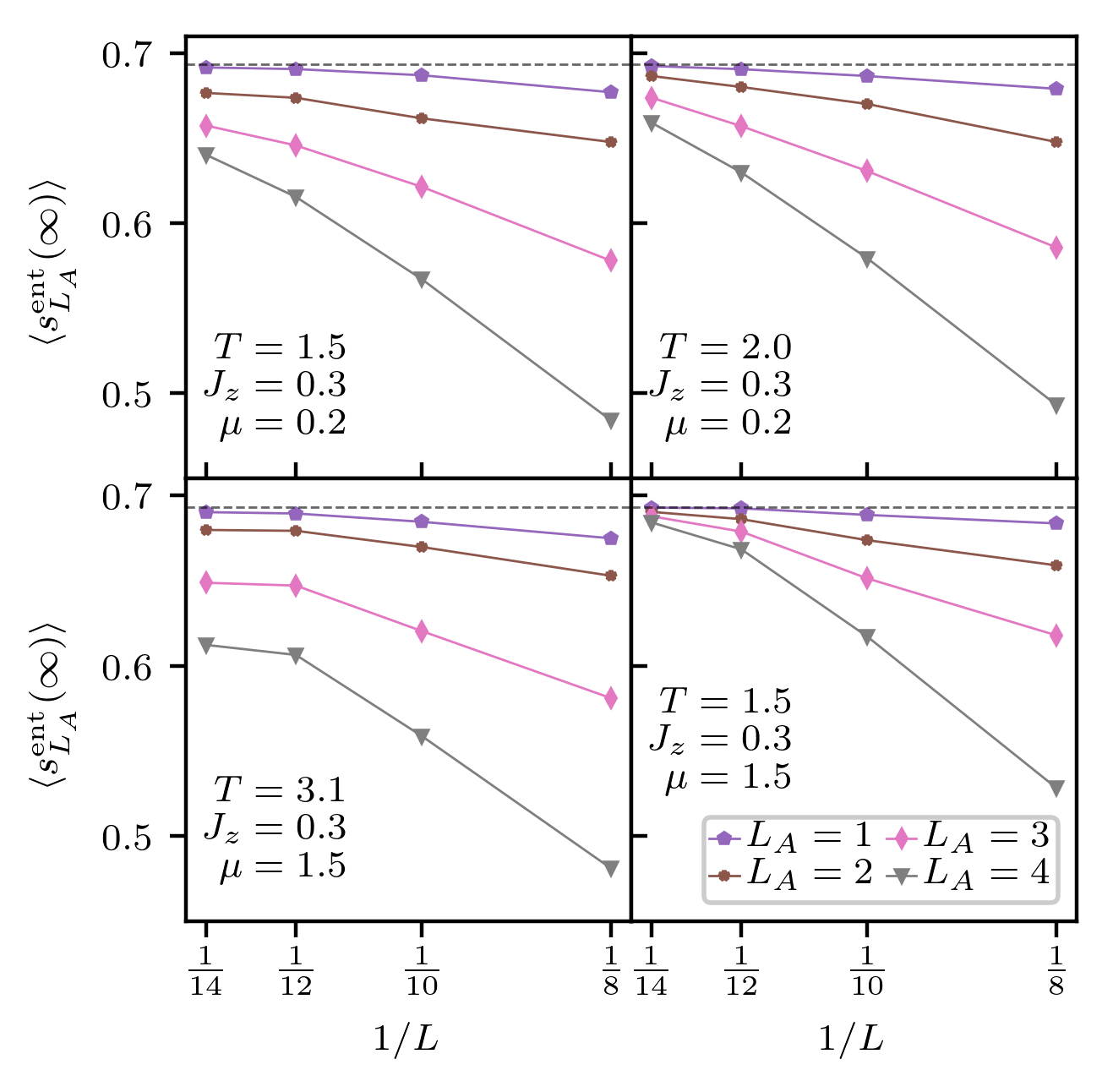}
  \caption{\label{finite1}
      Entanglement entropy density for different subsystem sizes \(L_A\) starting from the left
      end, and averaged over 100 time points from \(10^9-10^{12}\),
      long after the edge modes have melted. There is a clear trend towards the
      infinite temperature value of \(\log(2)\), dashed line, 
      as the chain length \(L\) is increased. 
  }
\end{figure}
Fig.~\ref{finite1} shows the very late time average of the
entanglement entropy per site as a function of inverse system size
$1/L$ for several subsystem sizes \(L_A\). We see that for the system
sizes accessible to us $s^{\rm ent}_{L_A}(\infty)$ approaches the infinite
temperature value \(\log(2)\). This is of course as expected, but it
allows us to quantify the role of finite-size effects.
In Fig.~\ref{finite3} we show the difference between the
entanglement entropy per site at finite times and the late time average
$s^{\rm ent}_{L_A}(\infty)$.
\begin{figure}[ht]
  \includegraphics[width=.49\textwidth]{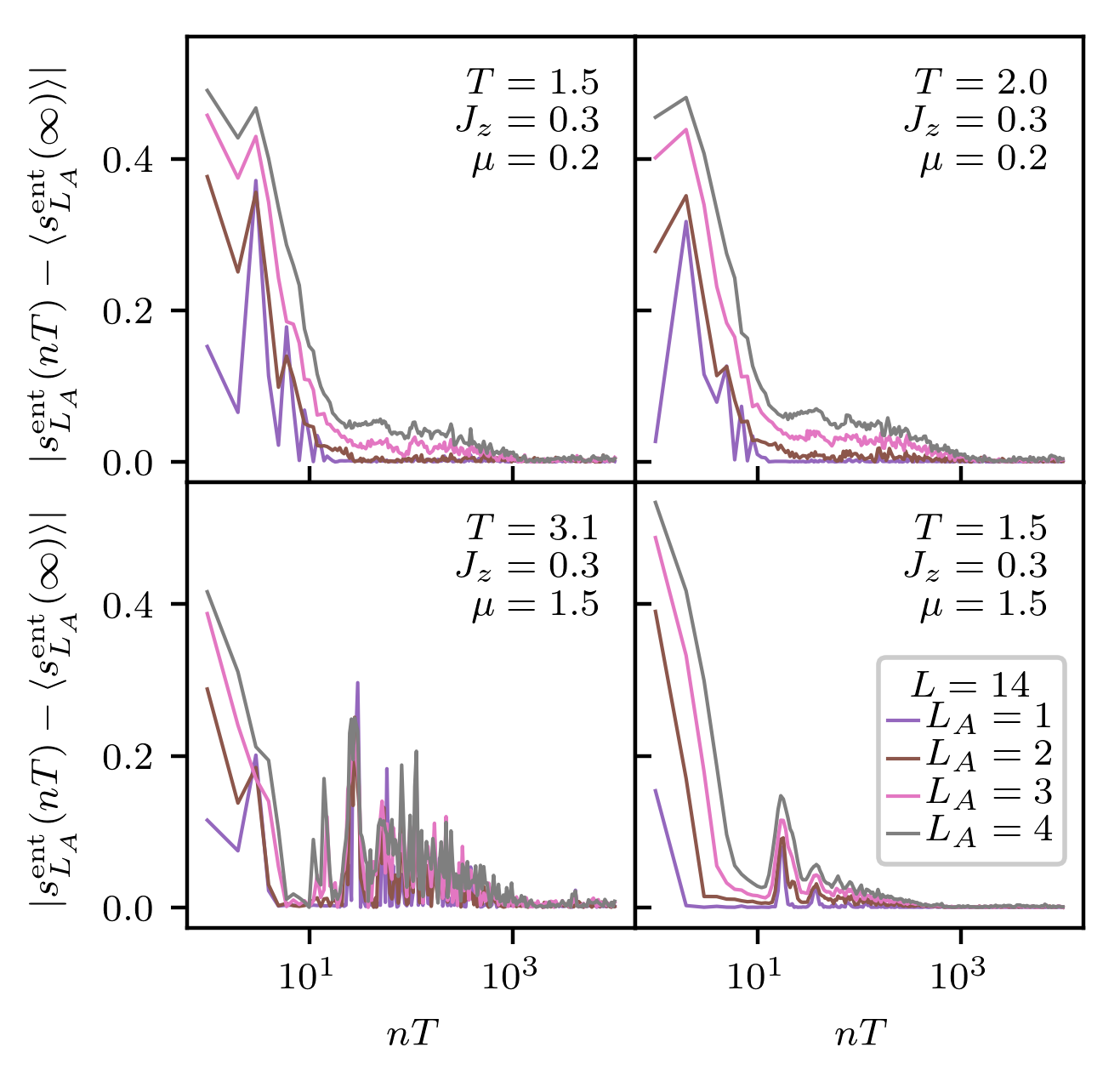}
  \caption{\label{finite3}
      Short time entanglement entropy density for different subsystem sizes \(L_A\),
      starting from the left end, for \(L = 14\). Plotted is the
      difference of the entanglement density from the very late entanglement density
      used in figure \ref{finite1}, where the latter is long 
      after the edge modes have melted. 
  }
\end{figure}
We see that $s^{\rm ent}_{L_A}(t)$ approaches its late time value, which
we have just argued to correspond to an infinite temperature state, on
time scales that are much shorter than the life times of the edge modes.
We note that reducing \(J_z\) will extend the lifetime of the edge
modes as discussed above, but not change the time scales shown in
Fig.~\ref{finite3}. In Fig.~\ref{finite2} we show the fluctuations of
\(\langle \sigma_{L/2}^z \rangle\) at late times. We see that as the
system size \(L\) is increased, fluctuations around the infinite
temperature value of zero are suppressed.
\begin{figure}[ht]
  \includegraphics[width=.49\textwidth]{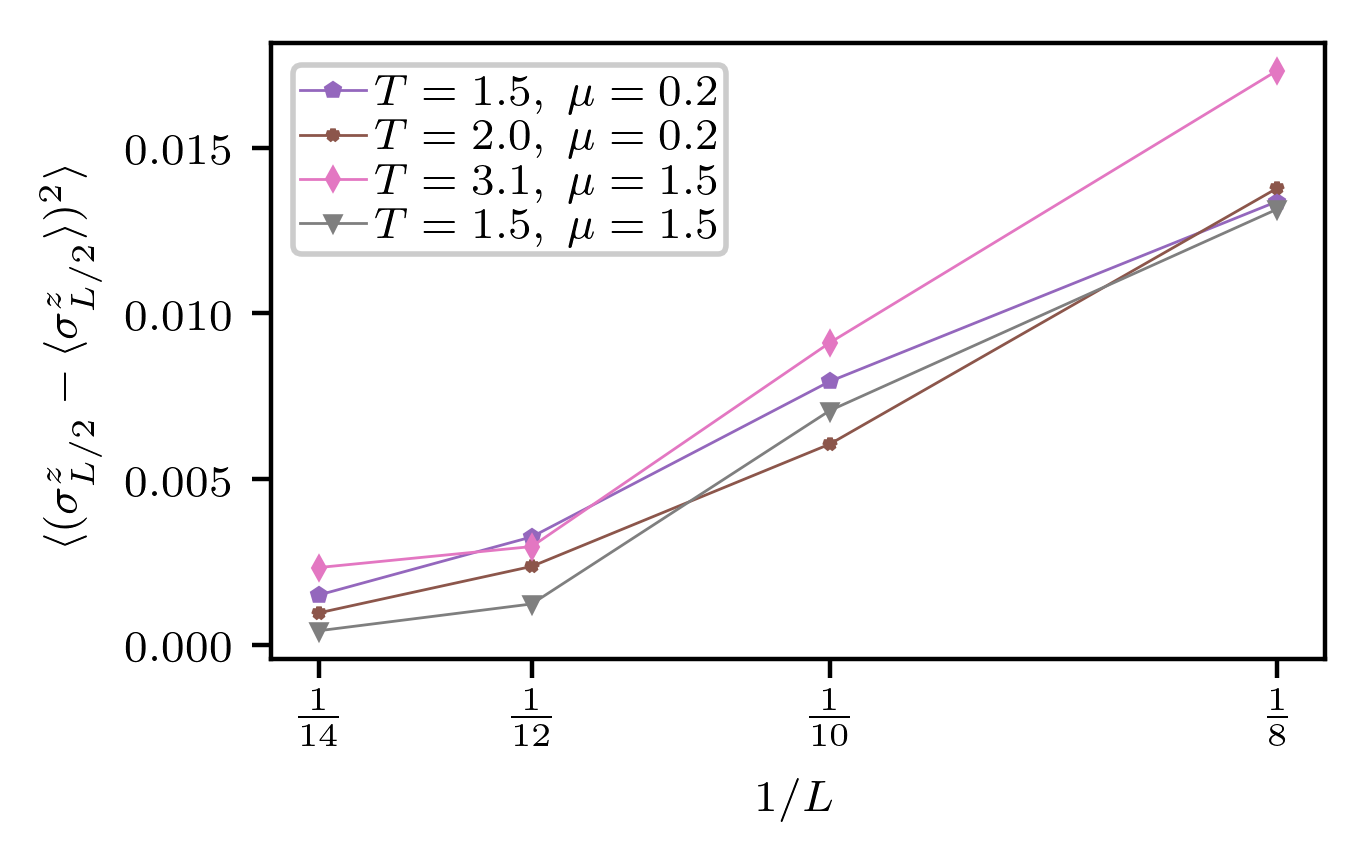}
  \caption{\label{finite2}
      Fluctuations of \(\langle \sigma_{L/2}^z \rangle\) after the initial
      decay. The average is taken over the last 40 time points visible in
      Figs.~\ref{fig3}, \ref{fig4}.
  }
\end{figure}

Inspection of Figs.~\ref{finite1}, \ref{finite2}, and \ref{finite3}
reveals that for parameters $T=3.1$, $J_z=0.3$, $\mu=1.5$, deviations
from the infinite temperature values are larger and convergence is
slower. This should be seen in the context that the lifetime of the
$\pi$ edge mode in this case has not yet saturated for system size
$L=14$, \emph{cf.} Fig.~\ref{fig4}. So while our finite-size analysis
is less conclusive in this case, our findings are compatible with the
general picture of local thermalization to an infinite temperature state
long before the edge modes start to decay.

\section{Floquet Hamiltonian} \label{FloquetH}
\begin{figure}[ht]
  \centering
  \includegraphics[width = .49\textwidth]{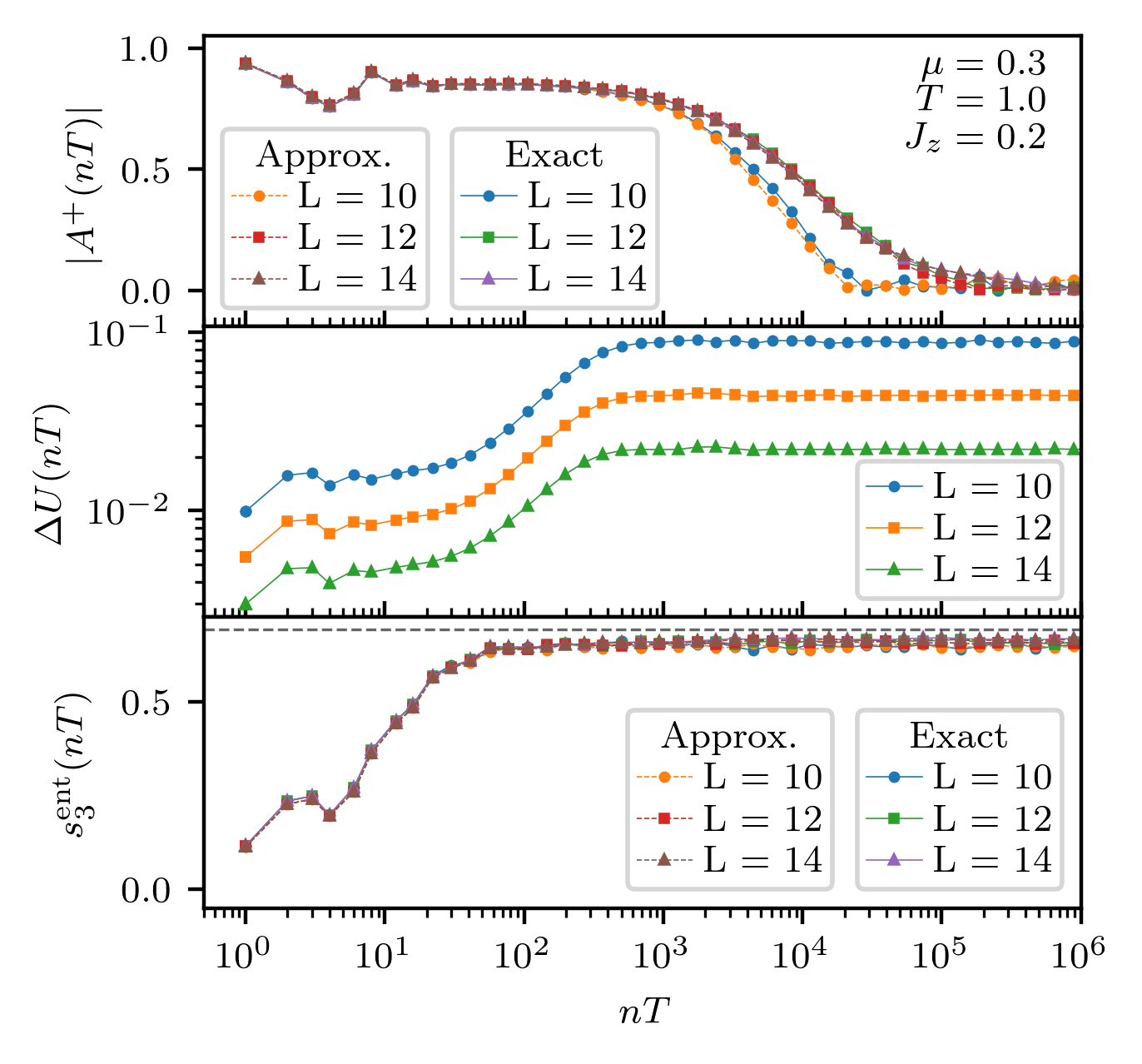}
  \caption{\label{fig6s_mzm} Dynamics of almost
    strong zero mode.
     Top row: \(A^+(nT)\) for the exact \(U\) and for the
    approximate \(U_{\rm approx}\) given by the leading order Magnus expansion.
    Middle row: $\Delta U (nT)$ as a function of stroboscopic time.
Bottom row: Time evolution of the entanglement
entropy density for a quench from a N\'eel state. The entanglement cut is
placed after the third site from the left end of the chain.
The onset of the decay for the almost strong zero modes is not tied to the short time
  features in \(\Delta U\) or \(s_3^{\rm ent}\). Lowering \(J_z\) to 0.1 (not shown) pushes
  the onset of the decay out to \(10^5\) periods while the steady states of the middle and
  bottom panels saturate before \(10^3\) and \(10^2\) respectively.
}
\end{figure}

It is instructive to investigate the existence of edge modes at the
level of the stroboscopic Floquet Hamiltonian obtained from
\(U(T)=e^{-iTH_F}\). We extract effective Floquet Hamiltonians around
two limits, high and low frequencies.
  As we are ultimately interested
in the behavior of large but finite systems, and short and intermediate
times, we set aside the issue of the convergence of such expansions.
In the small \(T\) limit of off-resonant driving, our Floquet Hamiltonian is
interacting and non-integrable
\be
H_F \approx H_F^{(0)}=\frac{1}{3} \left( J_{z}
H_{zz} + J_{x} H_{xx} + \mu H_{z} \right).
\ee
A quantitative measure of how well $H_F^{(0)}$ reproduces the time
evolution is provided by the normalized Frobenius norm of the difference
of evolution operators $\Delta(nT)=U(nT)-e^{-inTH_F^{(0)}}$
\be
\Delta U(nT)=\frac{1}{2^L}\sqrt{{\rm tr}\left[\big(\Delta(nT)\big)^\dagger\Delta(nT)\right]}.
\ee
Fig.~\ref{fig6s_mzm} shows that the dynamics induced by $H_F^{(0)}$ is in very
good agreement with the exact simulation for small values of $T$ (Appendix~\ref{appE}
discusses the choice of the lowest period). The existence of
almost strong edge modes in this setting, corresponds to the generalization of the
results of Kemp et al~\cite{Fendley17} to a quantum quench, for which
the system thermalizes on short, system-size independent time-scales,
while the zero mode persists over a much larger time-scale.

In the low frequency regime we can analyze the vicinity of the
exactly solvable limit $J_z=0$, $T\mu/3=\pi/2$ which supports a
strong $\pi$-mode. The Floquet Hamiltonian at this point is
\(TH_F^{(1)}=  \frac{TJ_x}{3} H_{xx} + \frac{\pi}{2} \mathcal{D}\),
and \(\sigma_1^x\) is an exact strong $\pi$-mode operator. Setting
\(\ J_zT/3 = \delta_{zz} \sim 0.082,\ T\mu/3 = \pi/2 + \delta_z
\sim \pi/2 -0.015\),
and \(J_x T/3 = \pi/4 + 0.1/3 = \theta_x\), we note that
we cannot perform a high-frequency expansion as \(TJ_{x}\) is not
small. Nevertheless, \(H_F\) to first order in
\(\delta_{zz},\delta_z\) but to arbitrary orders in \(T J_x\) may be
derived from an infinite resummation of the Baker-Campbell-Hausdorff
formula to obtain a non-local perturbed Ising model (see Appendix~\ref{appD}),
\begin{align}
  &TH_F^{(1)}
    \sim \frac{TJ_x}{3} H_{xx} + \frac{\pi}{2} \mathcal{D}
  + \delta_{zz} \left[  a_1 \left( h_{zyx}  + h_{xyz} \right) \right.\nonumber \\
  &+\left. a_2 h_{zz}^E + a_3 h_{zz}^B  + a_4 h_{xyyx}\right]
    + \delta_z \left[ b_1 h_z^E + b_2 h_z^B\right. \nonumber \\
  &\left.+ b_3 h_{xzx}  + b_4 \left( h_{xy} + h_{yx} \right)  \right].
\end{align}

\begin{figure}
  \centering
  \includegraphics[width = .49\textwidth]{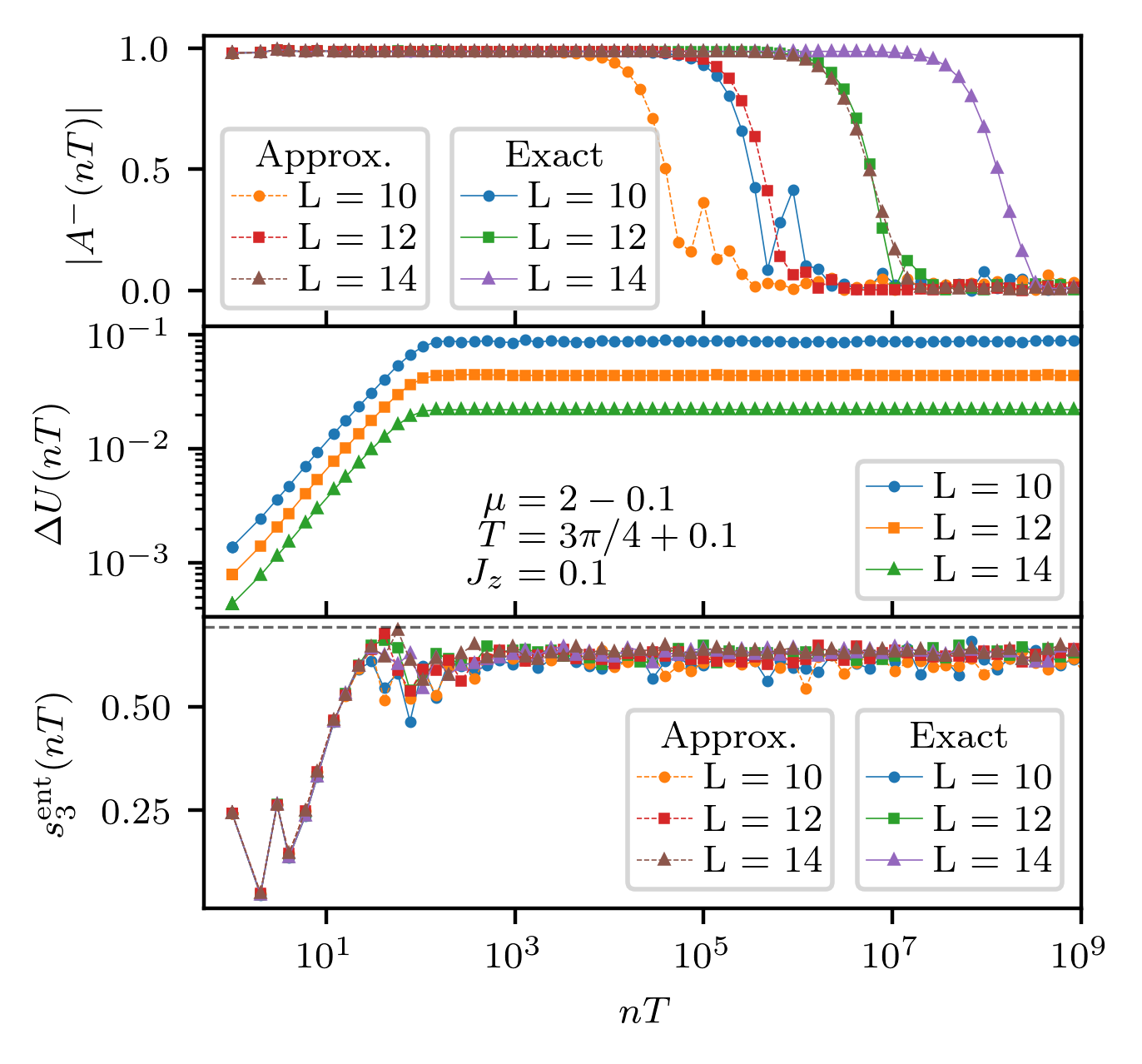}
  \caption{\label{fig6s_mpm} Top panel: Overlap \(A^-(nT)\) showing an almost
    strong \(\pi\) mode for both the exact time evolution and the approximate
    \(H_F\), the latter given by a BCH expansion about the point \(T\mu/3 =
    \pi/2 + \delta_z, TJ_x/3 = \pi/4 + \delta_T,T J_z/3 = \delta_{zz}\).
    Here, \(\delta_T = 0.1/3,\ \delta_z \sim -0.015,\ \delta_{zz} \sim 0.082.\)
    Middle
    panel: \(\Delta U(nT)\) as a function of stroboscopic time. Bottom panel:
    Time evolution of the entanglement entropy density with the entanglement cut
    placed after the third site from the left end of the chain. The initial
    state is a N{\'e}el state. While the expression for \(H_F\) fails to capture
    the long time dynamics accurately, it nevertheless manages to capture the
    exponential in system size dependence of the lifetime of the almost strong
    mode. We stress that the agreement of the exact $L=12$ and
    approximate $L=14$ results in the top panel is purely coincidental.}
\end{figure}

Here we have defined
\(h_{\alpha_1\dots\alpha_k}=\sum_j\sigma_j^{\alpha_1}\dots\sigma_{j+k-1}^{\alpha_k}
\equiv h^E_{\alpha_1\dots\alpha_k}+h^B_{\alpha_1\dots\alpha_k}\),
where $h^E$ denotes the contribution involving
the spins $\sigma_{1,L}^\alpha$ and $h^B$ the bulk part.
As expected, these additional terms still commute with
\(\mathcal{D}\).
Fig.~\ref{fig6s_mpm} shows that the dynamics of
\(A^-(nT)\) generated by this Hamiltonian
qualitatively agrees with the exact time evolution despite
\(\Delta U\) growing large at shorter times.
Taking into account
higher order corrections in \(\delta_{zz},\delta_z\)~\cite{Prosen18}
is expected to improve this agreement as long as we are close enough
to the exactly solvable point \(\delta_{zz},\delta_z\ll 1\).

\section{Conclusions} \label{conclu}
We have established the existence of long-lived
edge modes in periodically driven disorder-free systems with
interacting Floquet Hamiltonians. The lifetimes of these edge modes
are much longer than the time scales over which the system heats to
infinite temperature. This complements known results for edge modes in
periodically driven disordered and prethermal systems. The existence
of these modes imply robust edge states that survive heating, and open
up the possibility of using these states in quantum information and
computing.

Our work raises a number of questions. Most importantly one should understand
what determines the life times of the almost strong zero and $\pi$ modes. This
is currently under investigation. Another question is to what extent our
findings can be understood in terms of Ref.~\onlinecite{Abanin17b}
  where the authors give precise statements on the lifetime of prethermal
  physics for driven systems at high frequencies. In this
  paper, we avoided this regime due to the limits in system sizes accessible to
  us. To investigate this one should understand in what parameter regime
expansions of the Floquet Hamiltonian around solvable limits are asymptotic to
sufficiently high orders. It also would be interesting to explore eigenspectrum
phases with (almost) strong edge modes in spin-1 chains and higher dimensional
equilibrium as well as periodically driven systems. Other questions are whether
the strong edge modes in all free Floquet
SPTs~\cite{Kitagawa10,Zoller11,Benito14,Delplace14,Berdanier17,Yates17,Berdanier18,Yates18}
are equally robust to adding interactions. It is also interesting to explore the
connection between almost strong mode operators in interacting Floquet Hamiltonians and
edge modes of interacting topological phases~\cite{Fidk11,Verresen17}.

{\sl Acknowledgements:} We are grateful to Paul Fendley, Robert Konik,
Sid Parameswaran and Sthitadhi Roy for very helpful discussions. 
This work was supported by the US Department of Energy, Office of Science, 
Basic Energy Sciences, under Award No.~DE-SC0010821 (D.J.Y. and A.M.), and 
by the EPSRC under Grant No. EP/N01930X (F.H.L.E.). A.M. and F.H.L.E. also 
thank KITP for hospitality, which is supported by the National Science Foundation 
under Grant No. NSF PHY-1748958.



\appendix

\section{Explicit construction of the strong \(0,\pi\) mode operators
  for the binary drive}
\label{appA}
Our starting point is the time evolution of Majorana operators under
the two unitaries of our binary drive. Defining
\bea
U_1(T)&=&e^{-i \mu H_z \frac{T}{2}}\ ,\nn
U_2(T)&=&e^{-i J_x H_{xx} \frac{T}{2}}\ ,
\eea
we have
\be
U^\dagger_1(T)a_{2j}U_1(T)= \cos \left( T\mu \right)a_{2j}  + \sin
\left( T \mu \right)\ a_{2j-1}\ ,
\ee
\be
U^\dagger_1(T)a_{2j-1} U_1(T)
= \cos \left( T\mu \right)a_{2j-1}  - \sin \left( T \mu \right)
a_{2j}\ ,
\ee
\begin{align}
&U_2^\dagger(T) a_{2j} U_2(T)\nonumber\\
&=\begin{cases}
    a_{2j} & j = L\\
    \cos \left( TJ_x \right) a_{2j}  - \sin \left( T J_x \right) a_{2j+1} & j<L.
    \end{cases}\ ,
\end{align}
\begin{align}
&U_2^\dagger(T) a_{2j-1} U_2(T)\nonumber\\
  &=\begin{cases}
    a_{2j-1} & j = 1\\
    \cos \left( TJ_x \right) a_{2j-1}
+ \sin \left( T J_x \right) a_{2j-2} & j>1
    \end{cases}.
\end{align}
Denoting the time evolved Majorana operators by
$a_j(nT)=U^\dagger(nT)a_j U(nT)$, we can cast the evolution equations
in the form
\begin{align}
\begin{pmatrix} \vec{a}_{\rm odd}\big((n+1)T\big) \\ \vec{a}_{\rm even}\big((n+1)T\big) \end{pmatrix}
&\equiv M \begin{pmatrix} \vec{a}_{\rm odd}(nT) \\ \vec{a}_{\rm even}(nT) \end{pmatrix},
\end{align}
where $\vec{a}_{\rm odd}(nT)=(a_1(nT),a_3(nT),\ldots,a_{2L-1}(nT))$,
$\vec{a}_{\rm even}(nT)=(a_2(nT),a_4(nT),\ldots,a_{2L}(nT))$ and
\begin{align}
M&=      \left(
      \begin{array}{ccc|ccc}
        a'& & & c' & & \\
        b &a& & -d &c& \\
          &\ddots&\ddots & & \ddots & \ddots\\
        \hline
        -c & d & & a & b & \\
          & \ddots & \ddots & & \ddots & \ddots \\
         & & -c' & & & a'
      \end{array}
      \right)\ ,
\end{align}
and
\bea
a&=& \cos (T \mu) \cos (T J_x),\   b = \sin (T \mu) \sin (T J_x),\nn
c&=& -\sin (T \mu) \cos (T J_x),\   d = -\cos (T \mu) \sin (T J_x),\nn
a'&=& \cos (T \mu),\   c'= -\sin (T \mu).
\eea
We now use that the spin operators at the left edge of the chain have
a simple expression in terms of the Majorana fermions, i.e. \(\sigma^x_1(0)
= a_1(0)\). This suggests the following Ansatz for the zero and $\pi$
mode operators
\begin{align}
\Psi_\sigma=\sum_{j=1}^L \psi_j^{(\sigma)}
  a_{2j-1}+\phi_j^{(\sigma)} a_{2j}\ , 
\end{align}
where \(\phi_j^{(\sigma)}\) and \(\psi_j^{(\sigma)}\) are
respectively the amplitudes of the expansion for the even and odd
Majorana sublattices.  The requirement that
\be
U^\dagger(T)\Psi_\sigma U(T)=\cos(\sigma)\Psi_\sigma\ ,
\ee
translates into an eigenvalue equation for $M^T$
\begin{equation}
M^T
\begin{pmatrix}
\vec{\psi} \\
\vec{\phi}
\end{pmatrix}=\cos(\sigma)
\begin{pmatrix}
\vec{\psi} \\
\vec{\phi}
\end{pmatrix}.
\end{equation}
As $M^T$ is an orthogonal matrix we can equivalently consider the
eigenvalue equation for $M$, which we do in the following.
Denoting the $L\times L$ blocks of $M$ by
\begin{align}
  M = \begin{pmatrix}
    M_1 & M_2 \\ - {\mathbb{F}} M_2 {\mathbb{F}} &
    {\mathbb{F}} M_1 {\mathbb{F}}
  \end{pmatrix}\ ,\quad
{\mathbb{F}} = \begin{pmatrix} & & 1 \\ & 1 & \\ \reflectbox{$\ddots$} & &  \end{pmatrix},
\end{align}
we can block-diagonalize $M$ by
\be
\tilde{M}={\cal U}M{\cal U}^\dagger\ ,\quad
  {\cal U} = \frac{1}{\sqrt{2}}\begin{pmatrix}
    \mathbb{I} & i {\mathbb{F}} \\ \mathbb{I} & - i {\mathbb{F}}
  \end{pmatrix}.
\ee
This gives
\begin{equation}
\tilde{M}=\begin{pmatrix} \tilde{M}_1 & \\ & \tilde{M}_2 \end{pmatrix}
=\begin{pmatrix}
M_1 - i M_2 {\mathbb{F}} & \\ & M_1 + i M_2 {\mathbb{F}}
\end{pmatrix}.
\end{equation}
Finally we may diagonalize \(\tilde{M}_i = V_i \Lambda_i
V_i^\dagger\), where \(\Lambda_i\) is the diagonal matrix of eigenvalues
and the columns of \(V_i\) host the eigenvectors. Putting everything
together we can express $M$ in the form
\begin{align}
M    &=  \frac{1}{\sqrt{2}}
  \begin{pmatrix} V_1 & V_2\\
    -i {\mathbb{F}}V_1 & i {\mathbb{F}}V_2
  \end{pmatrix}
  \begin{pmatrix} \Lambda_1 & \\ & \Lambda_2 \end{pmatrix}
  \frac{1}{\sqrt{2}}
  \begin{pmatrix} V_1^\dagger & i V_1^\dagger {\mathbb{F}}\\
    V_2^\dagger & - i V_2^\dagger {\mathbb{F}}
  \end{pmatrix}\nonumber\\
  &\equiv W \Lambda W^\dagger\label{dyTrans}.
\end{align}
Since \(\tilde{M}_1^* = \tilde{M}_2\) and \(M\) is orthogonal, for
each eigenvector \(\ket{\lambda}\) of \(\tilde{M}_1\), there is an eigenvector
\(\ket{\lambda^*}\) of \(\tilde{M}_2\). For this reason, it suffices
to focus on \(\tilde{M}_1\). In the limit of large system size the
eigenvalue equation for $\tilde{M}_1$ turns into a
matrix recurrence relation of the form ($1\leq j<L/2$)
\begin{equation}
  \begin{pmatrix} b & id \\ -ic & a-\lambda \end{pmatrix}
  \begin{pmatrix} \psi_j \\ \phi_{L+1-j} \end{pmatrix}
  +
  \begin{pmatrix} a-\lambda & -ic \\ id & b \end{pmatrix}
  \begin{pmatrix} \psi_{j+1} \\ \phi_{L-j} \end{pmatrix}
  = 0,
\label{recurrence}
\end{equation}
while for $j=1$ we have
\be
\begin{pmatrix} a'-\lambda & -ic' \\ 0 & 0 \end{pmatrix}
\begin{pmatrix} \psi_1 \\ \phi_{L} \end{pmatrix}=0.
\label{recini}
\ee
Assuming that \(T\mu \ne \mathbb{Z} \pi,\ TJ_x\ne \mathbb{Z}\pi\) we
can rewrite this in the form
\be
  \begin{pmatrix} \psi_{j+1} \\ \phi_{L-j} \end{pmatrix}=C
  \begin{pmatrix} \psi_j \\ \phi_{L+1-j} \end{pmatrix},\quad j
\geq 1,
\label{nicerec}
\ee
where the matrix $C$ is
\begin{widetext}
\begin{align}
C  &=
    \frac{1}{\lambda \sin(T\mu) \sin(TJ_x)}
  \begin{pmatrix}
    \sin^2{\left (T \mu \right )} &
    -i \left[- \lambda \cos{\left (J_x T \right )} + \cos{\left (T \mu \right )}\right]\sin(T\mu) \\
     i\left[- \lambda \cos{\left (J_x T \right )} + \cos{\left (T \mu \right )}\right]\sin(T\mu) &
    1 - 2 \lambda \cos{\left (J_x T \right )} \cos{\left (T \mu \right )} + \cos^{2}{\left (T \mu \right )}.
\end{pmatrix}
\end{align}
\end{widetext}
The eigenvalues of $C$, for a given \(\lambda\) are
\begin{align}
\tilde{\epsilon}_{\pm}  &= \delta_{\lambda,+1} \left[ \cot(T\mu/2) \tan(TJ_x/2) \right]^{\pm 1}\nonumber\\
&- \delta_{\lambda,-1} \left[ \cot(T\mu/2) \cot(TJ_x/2) \right]^{\pm 1}.
\end{align}
The eigenvectors of \(C\) are,
\begin{align}
  \ket{\tilde{\epsilon}_+}
  &  = \begin{pmatrix} -i \sin(T\mu/2)\\ \cos (T\mu/2) \end{pmatrix},\
  \ket{\tilde{\epsilon}_-}
    = \begin{pmatrix} \cos(T\mu/2) \\ -i \sin(T\mu/2) \end{pmatrix},
\end{align}
and are independent of \(\lambda\).
The solutions to \fr{recini} for the relevant eigenvalues $\lambda=\pm
1$ are
\begin{align}
\begin{pmatrix} \psi_1 \\ \phi_L \end{pmatrix}
&= \begin{cases}
|\tilde{\epsilon}_-\rangle & \text{if }\lambda = 1\\
|\tilde{\epsilon}_+\rangle & \text{if }\lambda = -1
\end{cases}.
\end{align}
It is convenient to define,
\begin{align}
 \epsilon_\pm
  &= \mp\frac{\left[\cot(T\mu/2)\right]^{\pm 1}}{\tan(TJ_x/2)}.
\end{align}

Using the eigen-decomposition of \(C\) in \fr{nicerec},
we conclude that
\begin{subequations}
\begin{align}
  \begin{pmatrix}
    \psi_{n+1} \\
    \phi_{L-n}
  \end{pmatrix}
  &= \delta_{\lambda,1} \begin{pmatrix} \cos(T\mu/2) \\ -i \sin(T\mu/2) \end{pmatrix}
  \epsilon_-^n\nonumber\\
  &+ \delta_{\lambda,-1} \begin{pmatrix} -i \sin(T\mu/2) \\ \cos(T\mu/2) \end{pmatrix}
  \epsilon_+^n.\label{dyModes1}
\end{align}
\end{subequations}

So far we have neglected the fact that for $j=L/2$ the set of
recurrence relations is different. This is justified as long as
$|\epsilon_\pm|<1$ and $L\gg 1$. In this regime we can decompose the
zero and $\pi$ modes into their respective contributions centered on the
left and right edges respectively $\Psi_{0,\pi}\approx\Psi^L_{0,\pi}+\Psi^R_{0,\pi}$.
Focusing only on the left edge we have
\begin{align}
\Psi_{0} ^L &\approx \sum_{j \geq 1} \epsilon_-^{j-1}
  \biggl[\cos\left(\frac{T\mu}{2}\right) a_{2j-1}
-\sin\left(\frac{T\mu}{2}\right) a_{2j}\biggr],\nn
\Psi_{\pi}^L &\approx \sum_{j \geq 1}
\epsilon_+^{j-1}
  \biggl[ \sin\left(\frac{T\mu}{2}\right) a_{2j-1}
+ \cos\left( \frac{T\mu}{2}\right) a_{2j}\biggr].
\label{dyMPM1}
\end{align}
These are the expressions given in the main text.
\section{Edge mode diagnostic \(A(nT)\)}\label{appB}
\begin{figure}[ht]
  \centering
    \includegraphics[width = .49\textwidth]{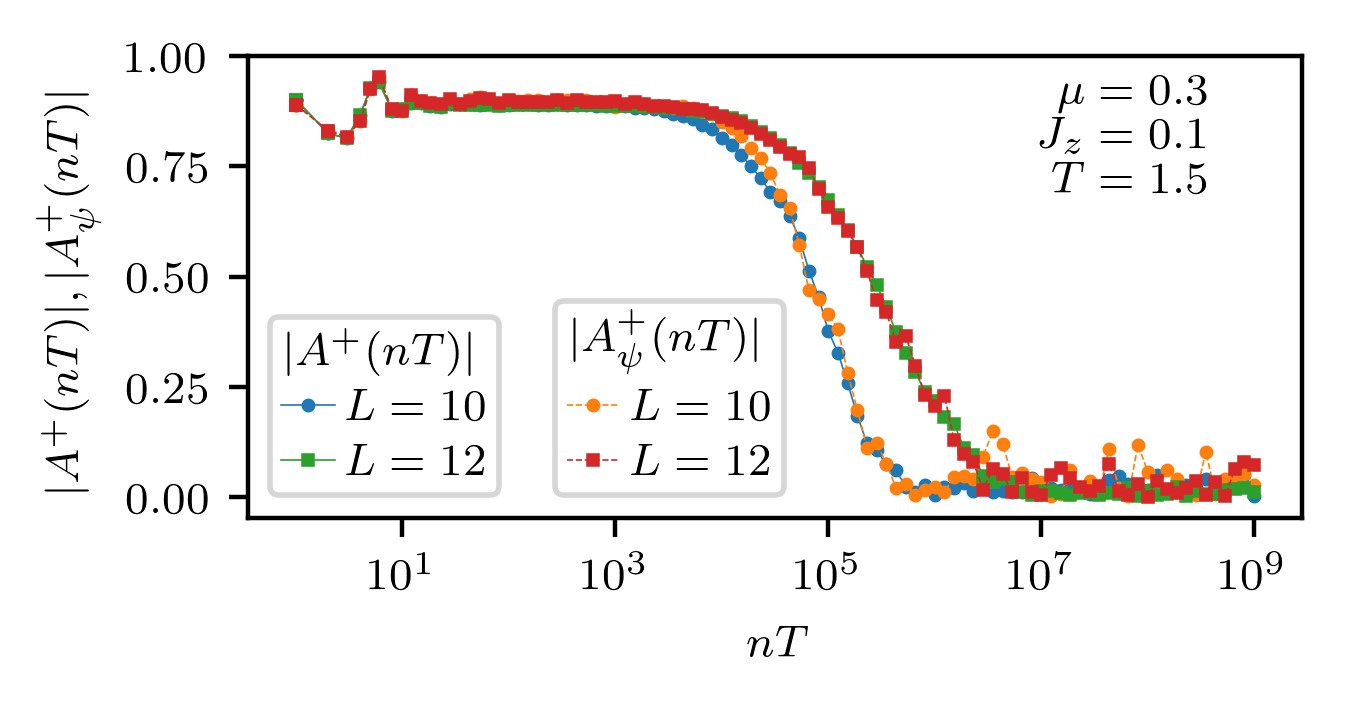}
    \caption{\label{avn} Symmetrized overlaps \(A^+(nT), A^+_{\psi}(nT)\), where
      \(|\psi\rangle\) is the N\'eel state. The time-evolution of the two
      quantities are almost identical, with both showing a lifetime that grows
      with system size \(L\), indicating an almost strong zero mode. }
\end{figure}

In this subsection we discuss the two measures \(A(nT), A_{\psi}(nT)\)
used to identify almost strong edge modes. The time evolution
operators commutes with rotations around the $z$-axis by 180 degrees,
and we therefore can choose the eigenstates of $U(T)$ to have definite
parity under these $\mathbb{Z}_2$ transformations
\bea
U(T)|m\rangle&=&e^{-iT\epsilon_m}|m\rangle\ ,\nn
{\cal D}|m\rangle&=&s_m|m\rangle\ ,\quad s_m=\pm.
\eea
Up to finite-size corrections exponentially small in system size a
strong zero mode \(\Psi_{0}\) sends eigenstates to eigenstates with
degenerate eigenvalues but with the opposite eigenvalue for \(\mathcal{D}\)
\bea
\Psi_0|m\rangle&\approx& |\bar{m}\rangle\ ,\quad
\epsilon_{\bar{m}}\approx\epsilon_m\ ,\nn
\label{paired}
\eea
A  strong \(\pi\) mode behaves similarly except that the quasi-energies
are shifted by \(\pi/T\). The spectral representation of $A(nT)$ reads
\begin{align}
  A(nT)
    &=\frac{1}{2^L}\sum_{m_1,m_2}
  \lvert \langle m_1| \sigma_1^x |m_2\rangle \rvert^2 e^{- i
    (\epsilon_{m_2} - \epsilon_{m_1})nT}\ .
\end{align}
As $\sigma^x_1$ is odd under the $\mathbb{Z}_2$ we have
\be
\sigma_1^x= c_0 \Psi_0 + c_\pi \Psi_\pi +\dots
\ee
The coefficients $c_0$ and $c_\pi$ are different from zero only if
strong zero/$\pi$ modes exist. Substituting this into the spectral
representation we have
\begin{subequations}
\begin{align}
A(nT)&=\frac{|c_0|^2}{2^{L}} \sum_{m_1,m_2} \lvert \langle m_1|\Psi_0|m_2\rangle \rvert^2 e^{-i(\epsilon_{m_2} - \epsilon_{m_1})nT}\nonumber\\
    &+\frac{|c_\pi|^2}{2^{L}}\sum_{m_1,m_2} \lvert \langle m_1|\Psi_\pi|m_2\rangle \rvert^2 e^{-i (\epsilon_{m_2} - \epsilon_{m_1})nT}\nonumber \\
&+ \frac{1}{2^{L}} \sum_{m_1,m_2}\biggl[c_0^*c_{\pi}\langle m_1| \Psi_0 |m_2\rangle\nn
      &\qquad\times \langle m_2| \Psi_\pi |m_1\rangle + {\rm
    h.c.}\biggr] e^{- i (\epsilon_{m_2} - \epsilon_{m_1})nT}+\dots
\label{AnT}
\end{align}
\end{subequations}
The exponential factors in \fr{AnT} will be strongly oscillating for
large $nT$ except for the $2^L$ terms in the double sums that correspond to
``paired'' states \fr{paired} and their $\pi$-mode analogues.
By the same arguments used in the thermalization context, the time
  average of the sum over oscillating terms becomes negligible at late
  times. Assuming that \(A(nT)\) relaxes, the oscillatory terms will therefore not
  contribute to the late-time behavior and 
\be
A(nT)\approx |c_0|^2 + |c_\pi|^2 e^{- i\pi n}.
\label{late}
\ee
The second measure we use is defined with respect to an initial state \(|\psi\rangle\)
\begin{align}
A_{\psi}(nT) = \langle\psi|U(nT)^{\dagger}\sigma^x_1U(nT)\sigma^x_1|\psi\rangle.
\end{align}
The physical meaning of this quantity is that we start from an initial
state \(|\psi\rangle\), flip a spin at site 1, then time-evolve until
time \(nT\), and flip the spin back again obtaining a state
\(\sigma^x_1 U(nT) \sigma^x_1|\psi\rangle\). \(A_{\psi}(nT)\) then
measures the overlap of this state with one where the initial state
was evolved up to time \(nT\) without the initial spin-flip \(U(nT) |\psi\rangle\).
Employing a spectral representation we have
\bea
A_{\psi}(nT)&=&\sum_{m_1,m_2}\langle m_1| \sigma_1^x
|m_2\rangle e^{- i (\epsilon_{m_2} - \epsilon_{m_1})nT}\nn
&&\qquad\times\ \langle\psi|m_1\rangle\langle m_2|\sigma^x_1|\psi\rangle.
\eea
Focussing on the non-oscillatory terms in this double sum (modulo
$(-1)^n$ in case of the $\pi$-mode) gives a late time contribution
that is the same as \fr{late}
\bea
A_{\psi}(nT)&\approx&|c_0|^2+|c_\pi|^2(-1)^n\ .
\eea
In Fig.~\ref{avn} we show the time-evolution of the symmetrized
autocorrelation functions \(A^+(nT), A^+_{\psi}(nT)\), and where
\(|\psi\rangle\) is chosen to be the N\'eel state. The figure shows
that an almost strong zero mode exists, and that the agreement between
the two measures is good.

\section{System size dependence of the phase diagram}\label{appC}

\begin{figure}
  \centering
  \includegraphics[width = .49\textwidth]{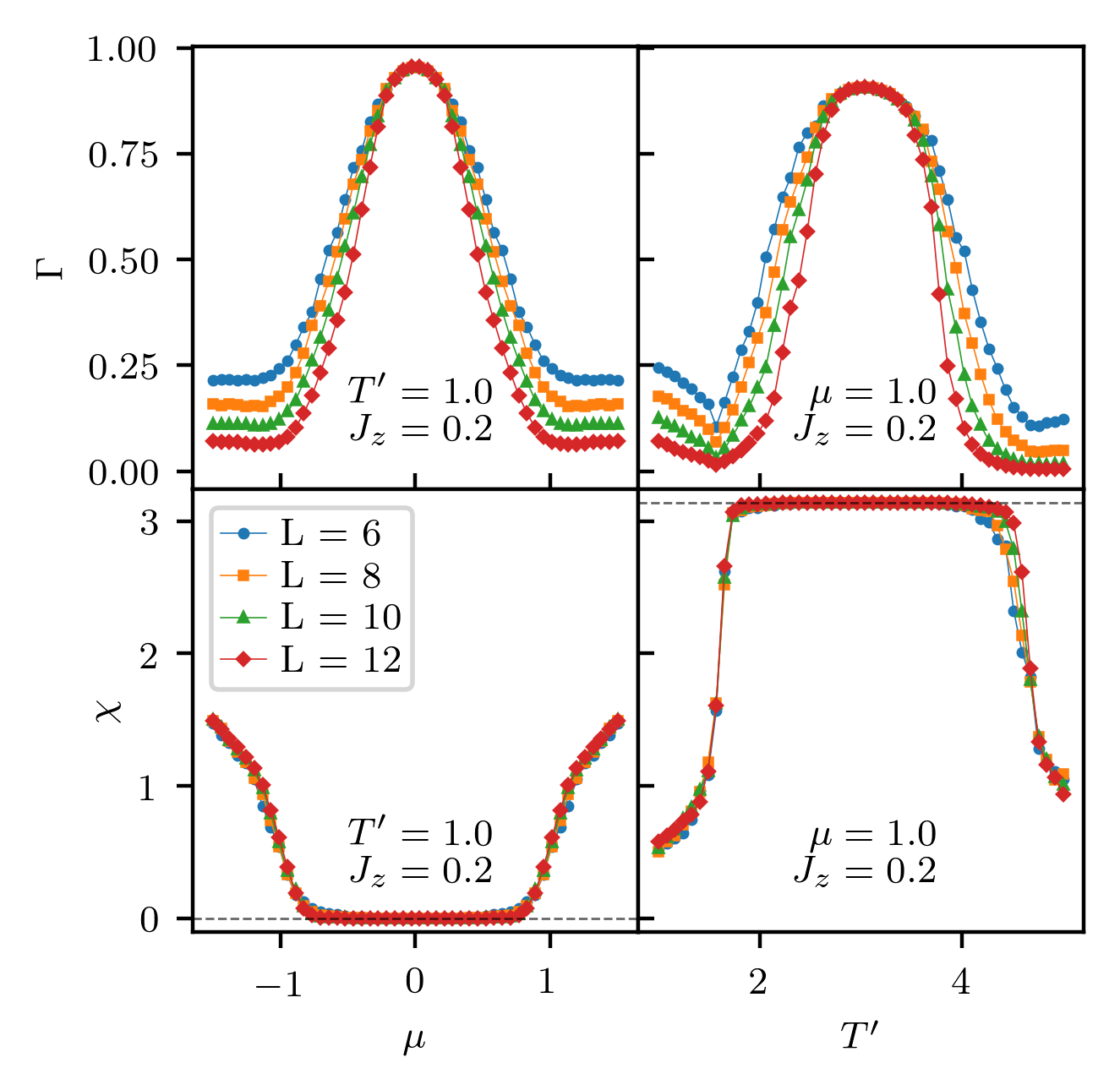}
  \caption{\label{fig4s} Top panels: Plots of \(\Gamma =\text{mean}_s
    \text{max}_{s'} |\langle s | \sigma_1^x | s' \rangle|^2\) where values of
    O(1) indicate an almost strong zero mode (left panel) or almost strong
    \(\pi\) mode (right panel). The nontrivial edge phases are robust to system
    size. Lower panels show another metric from directly measuring the pairing
    structure of the quasi-energy spectrum of the Floquet unitary.
    Lower left panel: Denoting \(|s'\rangle\) as the state that maximizes \(
    \lvert \langle s | \sigma_1^x | s' \rangle \rvert^2 \) for a given state
    \(|s\rangle\), \(\chi =\text{mean}_s \lvert \text{angle}\left(
      T\epsilon_s,T\epsilon_{s'}\right)\rvert \) for both almost strong
    \(0\) mode (left panel) and almost strong \(\pi\) mode (right panel).
    \(\text{angle}(x,y)\) finds the (smaller) angle between the two points
    on the unit circle. Note that while \(\Gamma\) cannot distinguish between \(0,\pi\) modes,
    \(\chi\) can.}
\end{figure}

Fig.~\ref{fig4s} plots two metrics for the almost strong modes \(\Gamma,\chi\),
each formally defined in the caption. \(\Gamma\) measures the extent to which
the operator \(\sigma^x_1\) connects different quasi-energy states, and does not
differentiate between whether these states have degenerate quasi-energies or not.

\(\chi\) measures the level of degeneracy for almost strong modes and/or the
level to which energies are separated by \(\pi/T\). The plots show that deep
within the phases, there is negligible system size dependence.

We have taken care to pick phases where only almost strong \(0\)
  or almost strong \(\pi\) mode exists, but \(\chi\) can also identify phases
  when both are present simultaneously. A flat plateau in \(\chi\) away from
  \(0, \pi\) would indicate the presence of coexisting \(0-\pi\) modes.

\section{Derivation of Floquet Hamiltonian with almost strong \(\pi\) mode}\label{appD}

We outline the derivation of \(H_F\), for the ternary drive, around the exactly
solvable limit \(J_z=0, T\mu/3=\pi/2\). Setting, \(\ J_zT/3 = \delta_{zz},\
T\mu/3 = \pi/2 + \delta_z\),
and \(J_x T/3 = \pi/4 + 0.1/3 = \theta_x\),
the Floquet unitary may be written as,
\begin{subequations}
\begin{align}
  U(T) &= e^{-i H_F T} = e^{-i \frac{TJ_z}{3}H_{zz}} e^{-i\frac{TJ_x}{3}H_{xx}} e^{-i \frac{T\mu}{3}H_{z}},\\
  &\equiv (-i)^{L-1}e^{-i\delta_{zz} H_{zz}} e^{-i\theta_x H_{xx}} e^{-i \delta_z H_{z}} e^{-i \frac{\pi}{2}\mathcal{D}},
\end{align}
\end{subequations}
where in the last line we have used that \(e^{-i\pi H_z/2}= (-i)^L \mathcal{D} =
(-i)^{L-1}e^{-i \frac{\pi}{2}\mathcal{D}}\). These steps are carried out to
explicitly show that we have a Floquet Hamiltonian that is non-local, and that
it is in fact the presence or absence of non-local term \(\mathcal{D}\) in the
Floquet Hamiltonian that determines whether \(\sigma^x_1\) is respectively a
strong \(\pi\) mode or a strong \(0\) mode.

In what, follows we will consider the limit of \(\delta_{zz},\delta_z \ll 1\). We
use the following formula from Baker-Campbell-Hausdorff (BCH),
\begin{equation}
  \log(\exp(X)\exp(Y)) = X + \frac{\rm{ad}_X e^{\rm{ad}_x}}{e^{\rm{ad}_X}-1}Y + \mathcal{O}(Y^2).
\end{equation}
Furthermore, we use the following identity,
\begin{equation}
  \frac{t}{1-e^{-t}} = \sum_{m = 0}^\infty \frac{B_{m}^+ t^m}{m!},
\end{equation}
where \(B_m^+\) are the Bernoulli numbers with \(B_1 = +\frac{1}{2}\). We first
note that \(\mathcal{D}\) commutes with everything, so it can be appended at the
end of the calculation. We first combine the exponentials containing \(\theta_x H_{xx}\) and
\(\delta_z H_z\), and define the resulting operator as \(Z_1\),
\begin{widetext}
\begin{subequations}
\begin{align}
  -iZ_1 &\sim -i\theta_x H_{xx}
          - i \delta_z \left( \sum_{n = 0}^\infty \frac{B_n^+ (-i \theta_x \text{ad}_{H_{xx}})^n}{n!} \right)H_{z}
+ \mathcal{O}(\delta_z^2),\\
&\sim -i\theta_x H_{xx} -i\delta_z \biggl\{ (\sigma_1^z + \sigma_L^z)\theta_x \cot(\theta_x)
+ \left(\frac{1 + 2 \theta_x \cot(2\theta_x)}{2} \right)  \sum_{i=2}^{L-1} \sigma_i^z\nonumber\\
&\qquad+ \left(\frac{-1 + 2 \theta_x \cot(2\theta_x)}{2}\right)
          \sum_{i=2}^{L-2} \sigma_{i-1}^x \sigma_i^z \sigma_{i+1}^x
- \theta_x \sum_{i=1}^{L-1} \left( \sigma_i^x\sigma_{i+1}^y + \sigma_{i}^y \sigma_{i+1}^x \right)
             \biggr\},\\
&\equiv -i\theta_x H_{xx} - i \delta_z \biggl\{  h_{z}^E \theta_x \cot(\theta_x)
+ h_{z}^B\left( \frac{1 + 2 \theta_x \cot(2\theta_x)}{2} \right)
+ h_{xzx}\left( \frac{-1+ 2 \theta_x \cot(2\theta_x)}{2} \right)
-\theta_x \left( h_{xy} + h_{yx} \right)\biggr\}.
\end{align}
\end{subequations}
\end{widetext}
Above we have used the notation
\(h_{\alpha_1\dots\alpha_k}=\sum_j\sigma_j^{\alpha_1}\dots\sigma_{j+k-1}^{\alpha_k}
\equiv h^E_{\alpha_1\dots\alpha_k}+h^B_{\alpha_1\dots\alpha_k}\), where $h^E$
denotes the contribution involving the spins $\sigma_{1,L}^\alpha$ and $h^B$ the
bulk part.

Next we combine the \(H_{zz}\) and \(Z_1\) exponentials using the same steps as
above, and only working to first order in \(\delta_z,\delta_{zz}\), obtain
the resulting operator \(Z_2\)
\begin{widetext}
\begin{subequations}
\begin{align}
  -i Z_2 & \sim - i Z_1 - i \delta_{zz} \biggl\{   \left( \sigma_1^z \sigma_2^z + \sigma_{L-1}^z \sigma_L^z \right)
           \theta_x \cot(\theta_x)
           + \left( \frac{1 + 2 \theta_x \cot(2\theta_x)}{2} \right)\sum_{i=2}^{L-2} \sigma_i^z\sigma_{i+1}^z\nonumber \\
         &\qquad - \left( \frac{-1 + 2 \theta_x \cot(2\theta_x)}{2} \right)
           \sum_{i=2}^{L-2} \left( \sigma_{i-1}^x \sigma_i^y \sigma_{i+1}^y \sigma_{i+2}^x \right)
           + \theta_x \sum_{i =2}^{L-1} \left( \sigma_{i-1}^z \sigma_i^y \sigma_{i+1}^x\
           + \sigma_{i-1}^x \sigma_i^y \sigma_{i+1}^z\right)
           \biggr\},\\
         &\equiv - i Z_1 - i \delta_{zz} \biggl\{h_{zz}^E \theta_x \cot(\theta_x)
           + h_{zz}^B\left( \frac{1 + 2 \theta_x \cot(2\theta_x)}{2} \right)
           - h_{xyyx}\left( \frac{-1 + 2 \theta_x \cot(2\theta_x)}{2} \right)
           + \theta_x \left( h_{zyx} + h_{xyz} \right)\biggr\}.
\end{align}
\end{subequations}
\end{widetext}
Now including the \(\pi\mathcal{D}/2\) term, we have our approximate \(H_F\),
\begin{widetext}
\begin{align}
  TH_F &\sim \theta_x H_{xx} + \frac{\pi}{2} \mathcal{D} + \delta_z \biggl\{  h_{z}^E \theta_x \cot(\theta_x)
          + h_{z}^B\left( \frac{1 + 2 \theta_x \cot(2\theta_x)}{2} \right)
          + h_{xzx}\left( \frac{-1+ 2 \theta_x \cot(2\theta_x)}{2} \right)
          -\theta_x \left( h_{xy} + h_{yx} \right)\biggr\}\nonumber\\
    &\qquad+ \delta_{zz} \biggl\{  h_{zz}^E \theta_x \cot(\theta_x)
           + h_{zz}^B\left( \frac{1 + 2 \theta_x \cot(2\theta_x)}{2} \right)
          - h_{xyyx}\left( \frac{-1 + 2 \theta_x \cot(2\theta_x)}{2} \right)
           + \theta_x \left( h_{zyx} + h_{xyz} \right)\biggr\}.
\end{align}
\end{widetext}
As we are working only to first order in \(\delta_{z}, \delta_{zz}\) we expect
this \(H_F\) to only be valid for short times.

While we have used the BCH formula above, a more systematic
approach following the methods in Ref.~\onlinecite{Prosen18} could prove useful for higher orders.
In fact we have checked that using the alternative approach of~\onlinecite{Prosen18}, and working to first order
gives the same form of \(H_F\).

\section{Discussion of parameters used}\label{appE}

\begin{figure}[ht]
  \centering
  \includegraphics[width = .49\textwidth]{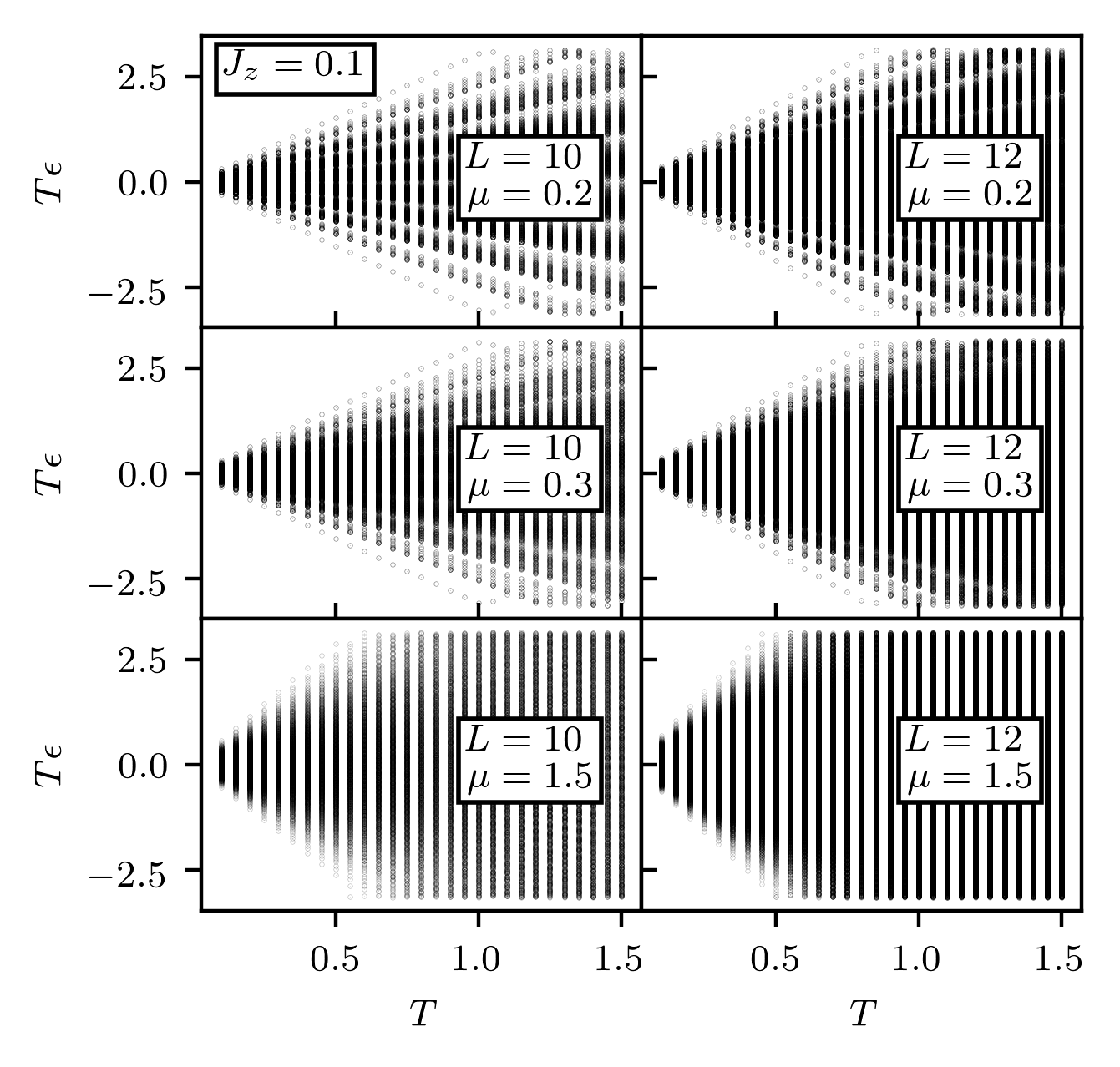}
  \caption{\label{fig11b}
    Quasi-energy spectrum plotted against \(T\) for select \(\mu\) values. For
    small system sizes, setting \(T\) too small can lead to the drive being
    off-resonant with the extensive many-body spectrum. We see that for \(L =
    12\), \(T = 1\) is a reasonable lower limit for our calculations.
  }
\end{figure}

Fig.~\ref{fig11b} shows how the many-particle quasi-energy spectrum evolves with system
size \(L\). For any \(T\), too small a system will not capture any true Floquet dynamics
as the spectrum will not reach the Floquet zone boundaries, and the system will always appear highly off-resonant.
For the parameters of our paper, \(T=1\) is a reasonable lower limit for the period. As a general rule, for a given \(T\), increasing
\(J_z\) or \(L\) or both, increases the number of many-body resonances.

\begin{thebibliography}{52}%
\makeatletter
\providecommand \@ifxundefined [1]{%
 \@ifx{#1\undefined}
}%
\providecommand \@ifnum [1]{%
 \ifnum #1\expandafter \@firstoftwo
 \else \expandafter \@secondoftwo
 \fi
}%
\providecommand \@ifx [1]{%
 \ifx #1\expandafter \@firstoftwo
 \else \expandafter \@secondoftwo
 \fi
}%
\providecommand \natexlab [1]{#1}%
\providecommand \enquote  [1]{``#1''}%
\providecommand \bibnamefont  [1]{#1}%
\providecommand \bibfnamefont [1]{#1}%
\providecommand \citenamefont [1]{#1}%
\providecommand \href@noop [0]{\@secondoftwo}%
\providecommand \href [0]{\begingroup \@sanitize@url \@href}%
\providecommand \@href[1]{\@@startlink{#1}\@@href}%
\providecommand \@@href[1]{\endgroup#1\@@endlink}%
\providecommand \@sanitize@url [0]{\catcode `\\12\catcode `\$12\catcode
  `\&12\catcode `\#12\catcode `\^12\catcode `\_12\catcode `\%12\relax}%
\providecommand \@@startlink[1]{}%
\providecommand \@@endlink[0]{}%
\providecommand \url  [0]{\begingroup\@sanitize@url \@url }%
\providecommand \@url [1]{\endgroup\@href {#1}{\urlprefix }}%
\providecommand \urlprefix  [0]{URL }%
\providecommand \Eprint [0]{\href }%
\providecommand \doibase [0]{http://dx.doi.org/}%
\providecommand \selectlanguage [0]{\@gobble}%
\providecommand \bibinfo  [0]{\@secondoftwo}%
\providecommand \bibfield  [0]{\@secondoftwo}%
\providecommand \translation [1]{[#1]}%
\providecommand \BibitemOpen [0]{}%
\providecommand \bibitemStop [0]{}%
\providecommand \bibitemNoStop [0]{.\EOS\space}%
\providecommand \EOS [0]{\spacefactor3000\relax}%
\providecommand \BibitemShut  [1]{\csname bibitem#1\endcsname}%
\let\auto@bib@innerbib\@empty
\bibitem [{\citenamefont {Kitaev}(2006)}]{Kitaev06}%
  \BibitemOpen
  \bibfield  {author} {\bibinfo {author} {\bibfnamefont {A.}~\bibnamefont
  {Kitaev}},\ }\href {\doibase https://doi.org/10.1016/j.aop.2005.10.005}
  {\bibfield  {journal} {\bibinfo  {journal} {Annals of Physics}\ }\textbf
  {\bibinfo {volume} {321}},\ \bibinfo {pages} {2 } (\bibinfo {year} {2006})},\
  \bibinfo {note} {january Special Issue}\BibitemShut {NoStop}%
\bibitem [{\citenamefont {Nayak}\ \emph {et~al.}(2008)\citenamefont {Nayak},
  \citenamefont {Simon}, \citenamefont {Stern}, \citenamefont {Freedman},\ and\
  \citenamefont {Das~Sarma}}]{NayakRMP08}%
  \BibitemOpen
  \bibfield  {author} {\bibinfo {author} {\bibfnamefont {C.}~\bibnamefont
  {Nayak}}, \bibinfo {author} {\bibfnamefont {S.~H.}\ \bibnamefont {Simon}},
  \bibinfo {author} {\bibfnamefont {A.}~\bibnamefont {Stern}}, \bibinfo
  {author} {\bibfnamefont {M.}~\bibnamefont {Freedman}}, \ and\ \bibinfo
  {author} {\bibfnamefont {S.}~\bibnamefont {Das~Sarma}},\ }\href {\doibase
  10.1103/RevModPhys.80.1083} {\bibfield  {journal} {\bibinfo  {journal} {Rev.
  Mod. Phys.}\ }\textbf {\bibinfo {volume} {80}},\ \bibinfo {pages} {1083}
  (\bibinfo {year} {2008})}\BibitemShut {NoStop}%
\bibitem [{\citenamefont {Fendley}\ \emph {et~al.}(2009)\citenamefont
  {Fendley}, \citenamefont {Fisher},\ and\ \citenamefont {Nayak}}]{Fendley09}%
  \BibitemOpen
  \bibfield  {author} {\bibinfo {author} {\bibfnamefont {P.}~\bibnamefont
  {Fendley}}, \bibinfo {author} {\bibfnamefont {M.~P.}\ \bibnamefont {Fisher}},
  \ and\ \bibinfo {author} {\bibfnamefont {C.}~\bibnamefont {Nayak}},\ }\href
  {\doibase https://doi.org/10.1016/j.aop.2009.03.005} {\bibfield  {journal}
  {\bibinfo  {journal} {Annals of Physics}\ }\textbf {\bibinfo {volume}
  {324}},\ \bibinfo {pages} {1547 } (\bibinfo {year} {2009})},\ \bibinfo {note}
  {july 2009 Special Issue}\BibitemShut {NoStop}%
\bibitem [{\citenamefont {Alicea}(2012)}]{Alicea12}%
  \BibitemOpen
  \bibfield  {author} {\bibinfo {author} {\bibfnamefont {J.}~\bibnamefont
  {Alicea}},\ }\href {\doibase 10.1088/0034-4885/75/7/076501} {\bibfield
  {journal} {\bibinfo  {journal} {Reports on Progress in Physics}\ }\textbf
  {\bibinfo {volume} {75}},\ \bibinfo {pages} {076501} (\bibinfo {year}
  {2012})}\BibitemShut {NoStop}%
\bibitem [{\citenamefont {Beenakker}(2013)}]{Beenaker13}%
  \BibitemOpen
  \bibfield  {author} {\bibinfo {author} {\bibfnamefont {C.}~\bibnamefont
  {Beenakker}},\ }\href {\doibase 10.1146/annurev-conmatphys-030212-184337}
  {\bibfield  {journal} {\bibinfo  {journal} {Annual Review of Condensed Matter
  Physics}\ }\textbf {\bibinfo {volume} {4}},\ \bibinfo {pages} {113} (\bibinfo
  {year} {2013})}\BibitemShut {NoStop}%
\bibitem [{\citenamefont {Fendley}(2016)}]{Fendley16}%
  \BibitemOpen
  \bibfield  {author} {\bibinfo {author} {\bibfnamefont {P.}~\bibnamefont
  {Fendley}},\ }\href {http://stacks.iop.org/1751-8121/49/i=30/a=30LT01}
  {\bibfield  {journal} {\bibinfo  {journal} {Journal of Physics A:
  Mathematical and Theoretical}\ }\textbf {\bibinfo {volume} {49}},\ \bibinfo
  {pages} {30LT01} (\bibinfo {year} {2016})}\BibitemShut {NoStop}%
\bibitem [{\citenamefont {Kemp}\ \emph {et~al.}(2017)\citenamefont {Kemp},
  \citenamefont {Yao}, \citenamefont {Laumann},\ and\ \citenamefont
  {Fendley}}]{Fendley17}%
  \BibitemOpen
  \bibfield  {author} {\bibinfo {author} {\bibfnamefont {J.}~\bibnamefont
  {Kemp}}, \bibinfo {author} {\bibfnamefont {N.~Y.}\ \bibnamefont {Yao}},
  \bibinfo {author} {\bibfnamefont {C.~R.}\ \bibnamefont {Laumann}}, \ and\
  \bibinfo {author} {\bibfnamefont {P.}~\bibnamefont {Fendley}},\ }\href
  {http://stacks.iop.org/1742-5468/2017/i=6/a=063105} {\bibfield  {journal}
  {\bibinfo  {journal} {Journal of Statistical Mechanics: Theory and
  Experiment}\ }\textbf {\bibinfo {volume} {2017}},\ \bibinfo {pages} {063105}
  (\bibinfo {year} {2017})}\BibitemShut {NoStop}%
\bibitem [{\citenamefont {Else}\ \emph
  {et~al.}(2017{\natexlab{a}})\citenamefont {Else}, \citenamefont {Fendley},
  \citenamefont {Kemp},\ and\ \citenamefont {Nayak}}]{Nayak17}%
  \BibitemOpen
  \bibfield  {author} {\bibinfo {author} {\bibfnamefont {D.~V.}\ \bibnamefont
  {Else}}, \bibinfo {author} {\bibfnamefont {P.}~\bibnamefont {Fendley}},
  \bibinfo {author} {\bibfnamefont {J.}~\bibnamefont {Kemp}}, \ and\ \bibinfo
  {author} {\bibfnamefont {C.}~\bibnamefont {Nayak}},\ }\href {\doibase
  10.1103/PhysRevX.7.041062} {\bibfield  {journal} {\bibinfo  {journal} {Phys.
  Rev. X}\ }\textbf {\bibinfo {volume} {7}},\ \bibinfo {pages} {041062}
  (\bibinfo {year} {2017}{\natexlab{a}})}\BibitemShut {NoStop}%
\bibitem [{\citenamefont {Vasiloiu}\ \emph {et~al.}(2018)\citenamefont
  {Vasiloiu}, \citenamefont {Carollo},\ and\ \citenamefont
  {Garrahan}}]{Garrahan18}%
  \BibitemOpen
  \bibfield  {author} {\bibinfo {author} {\bibfnamefont {L.~M.}\ \bibnamefont
  {Vasiloiu}}, \bibinfo {author} {\bibfnamefont {F.}~\bibnamefont {Carollo}}, \
  and\ \bibinfo {author} {\bibfnamefont {J.~P.}\ \bibnamefont {Garrahan}},\
  }\href {\doibase 10.1103/PhysRevB.98.094308} {\bibfield  {journal} {\bibinfo
  {journal} {Phys. Rev. B}\ }\textbf {\bibinfo {volume} {98}},\ \bibinfo
  {pages} {094308} (\bibinfo {year} {2018})}\BibitemShut {NoStop}%
\bibitem [{\citenamefont {Vasiloiu}\ \emph {et~al.}(shed)\citenamefont
  {Vasiloiu}, \citenamefont {Carollo}, \citenamefont {Marcuzzi},\ and\
  \citenamefont {Garrahan}}]{Garrahan19}%
  \BibitemOpen
  \bibfield  {author} {\bibinfo {author} {\bibfnamefont {L.~M.}\ \bibnamefont
  {Vasiloiu}}, \bibinfo {author} {\bibfnamefont {F.}~\bibnamefont {Carollo}},
  \bibinfo {author} {\bibfnamefont {M.}~\bibnamefont {Marcuzzi}}, \ and\
  \bibinfo {author} {\bibfnamefont {J.~P.}\ \bibnamefont {Garrahan}},\
  }\href@noop {} {\bibfield  {journal} {\bibinfo  {journal} {arXiv:1901.10211}\
  } (\bibinfo {year} {unpublished})}\BibitemShut {NoStop}%
\bibitem [{\citenamefont {Thakurathi}\ \emph {et~al.}(2013)\citenamefont
  {Thakurathi}, \citenamefont {Patel}, \citenamefont {Sen},\ and\ \citenamefont
  {Dutta}}]{Sen13}%
  \BibitemOpen
  \bibfield  {author} {\bibinfo {author} {\bibfnamefont {M.}~\bibnamefont
  {Thakurathi}}, \bibinfo {author} {\bibfnamefont {A.~A.}\ \bibnamefont
  {Patel}}, \bibinfo {author} {\bibfnamefont {D.}~\bibnamefont {Sen}}, \ and\
  \bibinfo {author} {\bibfnamefont {A.}~\bibnamefont {Dutta}},\ }\href
  {\doibase 10.1103/PhysRevB.88.155133} {\bibfield  {journal} {\bibinfo
  {journal} {Phys. Rev. B}\ }\textbf {\bibinfo {volume} {88}},\ \bibinfo
  {pages} {155133} (\bibinfo {year} {2013})}\BibitemShut {NoStop}%
\bibitem [{\citenamefont {Bahri}\ \emph {et~al.}(2015)\citenamefont {Bahri},
  \citenamefont {Ronen}, \citenamefont {Altman},\ and\ \citenamefont
  {Vishwanath}}]{Bahri15}%
  \BibitemOpen
  \bibfield  {author} {\bibinfo {author} {\bibfnamefont {Y.}~\bibnamefont
  {Bahri}}, \bibinfo {author} {\bibfnamefont {R.}~\bibnamefont {Ronen}},
  \bibinfo {author} {\bibfnamefont {E.}~\bibnamefont {Altman}}, \ and\ \bibinfo
  {author} {\bibfnamefont {A.}~\bibnamefont {Vishwanath}},\ }\href@noop {}
  {\bibfield  {journal} {\bibinfo  {journal} {Nature Communications}\ }\textbf
  {\bibinfo {volume} {6}},\ \bibinfo {pages} {7341} (\bibinfo {year}
  {2015})}\BibitemShut {NoStop}%
\bibitem [{\citenamefont {Khemani}\ \emph {et~al.}(2016)\citenamefont
  {Khemani}, \citenamefont {Lazarides}, \citenamefont {Moessner},\ and\
  \citenamefont {Sondhi}}]{Khemani16}%
  \BibitemOpen
  \bibfield  {author} {\bibinfo {author} {\bibfnamefont {V.}~\bibnamefont
  {Khemani}}, \bibinfo {author} {\bibfnamefont {A.}~\bibnamefont {Lazarides}},
  \bibinfo {author} {\bibfnamefont {R.}~\bibnamefont {Moessner}}, \ and\
  \bibinfo {author} {\bibfnamefont {S.~L.}\ \bibnamefont {Sondhi}},\ }\href
  {\doibase 10.1103/PhysRevLett.116.250401} {\bibfield  {journal} {\bibinfo
  {journal} {Phys. Rev. Lett.}\ }\textbf {\bibinfo {volume} {116}},\ \bibinfo
  {pages} {250401} (\bibinfo {year} {2016})}\BibitemShut {NoStop}%
\bibitem [{\citenamefont {Sreejith}\ \emph {et~al.}(2016)\citenamefont
  {Sreejith}, \citenamefont {Lazarides},\ and\ \citenamefont
  {Moessner}}]{Para16}%
  \BibitemOpen
  \bibfield  {author} {\bibinfo {author} {\bibfnamefont {G.~J.}\ \bibnamefont
  {Sreejith}}, \bibinfo {author} {\bibfnamefont {A.}~\bibnamefont {Lazarides}},
  \ and\ \bibinfo {author} {\bibfnamefont {R.}~\bibnamefont {Moessner}},\
  }\href {\doibase 10.1103/PhysRevB.94.045127} {\bibfield  {journal} {\bibinfo
  {journal} {Phys. Rev. B}\ }\textbf {\bibinfo {volume} {94}},\ \bibinfo
  {pages} {045127} (\bibinfo {year} {2016})}\BibitemShut {NoStop}%
\bibitem [{\citenamefont {Potirniche}\ \emph {et~al.}(2017)\citenamefont
  {Potirniche}, \citenamefont {Potter}, \citenamefont {Schleier-Smith},
  \citenamefont {Vishwanath},\ and\ \citenamefont {Yao}}]{Yao17}%
  \BibitemOpen
  \bibfield  {author} {\bibinfo {author} {\bibfnamefont {I.-D.}\ \bibnamefont
  {Potirniche}}, \bibinfo {author} {\bibfnamefont {A.~C.}\ \bibnamefont
  {Potter}}, \bibinfo {author} {\bibfnamefont {M.}~\bibnamefont
  {Schleier-Smith}}, \bibinfo {author} {\bibfnamefont {A.}~\bibnamefont
  {Vishwanath}}, \ and\ \bibinfo {author} {\bibfnamefont {N.~Y.}\ \bibnamefont
  {Yao}},\ }\href {\doibase 10.1103/PhysRevLett.119.123601} {\bibfield
  {journal} {\bibinfo  {journal} {Phys. Rev. Lett.}\ }\textbf {\bibinfo
  {volume} {119}},\ \bibinfo {pages} {123601} (\bibinfo {year}
  {2017})}\BibitemShut {NoStop}%
\bibitem [{\citenamefont {Kumar}\ \emph {et~al.}(2018)\citenamefont {Kumar},
  \citenamefont {Dumitrescu},\ and\ \citenamefont {Potter}}]{Potter18}%
  \BibitemOpen
  \bibfield  {author} {\bibinfo {author} {\bibfnamefont {A.}~\bibnamefont
  {Kumar}}, \bibinfo {author} {\bibfnamefont {P.~T.}\ \bibnamefont
  {Dumitrescu}}, \ and\ \bibinfo {author} {\bibfnamefont {A.~C.}\ \bibnamefont
  {Potter}},\ }\href {\doibase 10.1103/PhysRevB.97.224302} {\bibfield
  {journal} {\bibinfo  {journal} {Phys. Rev. B}\ }\textbf {\bibinfo {volume}
  {97}},\ \bibinfo {pages} {224302} (\bibinfo {year} {2018})}\BibitemShut
  {NoStop}%
\bibitem [{\citenamefont {Wen}(2017)}]{Wen17}%
  \BibitemOpen
  \bibfield  {author} {\bibinfo {author} {\bibfnamefont {X.-G.}\ \bibnamefont
  {Wen}},\ }\href {\doibase 10.1103/RevModPhys.89.041004} {\bibfield  {journal}
  {\bibinfo  {journal} {Rev. Mod. Phys.}\ }\textbf {\bibinfo {volume} {89}},\
  \bibinfo {pages} {041004} (\bibinfo {year} {2017})}\BibitemShut {NoStop}%
\bibitem [{\citenamefont {Iadecola}\ \emph {et~al.}(2015)\citenamefont
  {Iadecola}, \citenamefont {Santos},\ and\ \citenamefont {Chamon}}]{Chamon15}%
  \BibitemOpen
  \bibfield  {author} {\bibinfo {author} {\bibfnamefont {T.}~\bibnamefont
  {Iadecola}}, \bibinfo {author} {\bibfnamefont {L.~H.}\ \bibnamefont
  {Santos}}, \ and\ \bibinfo {author} {\bibfnamefont {C.}~\bibnamefont
  {Chamon}},\ }\href {\doibase 10.1103/PhysRevB.92.125107} {\bibfield
  {journal} {\bibinfo  {journal} {Phys. Rev. B}\ }\textbf {\bibinfo {volume}
  {92}},\ \bibinfo {pages} {125107} (\bibinfo {year} {2015})}\BibitemShut
  {NoStop}%
\bibitem [{\citenamefont {Kitagawa}\ \emph {et~al.}(2010)\citenamefont
  {Kitagawa}, \citenamefont {Berg}, \citenamefont {Rudner},\ and\ \citenamefont
  {Demler}}]{Kitagawa10}%
  \BibitemOpen
  \bibfield  {author} {\bibinfo {author} {\bibfnamefont {T.}~\bibnamefont
  {Kitagawa}}, \bibinfo {author} {\bibfnamefont {E.}~\bibnamefont {Berg}},
  \bibinfo {author} {\bibfnamefont {M.}~\bibnamefont {Rudner}}, \ and\ \bibinfo
  {author} {\bibfnamefont {E.}~\bibnamefont {Demler}},\ }\href {\doibase
  10.1103/PhysRevB.82.235114} {\bibfield  {journal} {\bibinfo  {journal} {Phys.
  Rev. B}\ }\textbf {\bibinfo {volume} {82}},\ \bibinfo {pages} {235114}
  (\bibinfo {year} {2010})}\BibitemShut {NoStop}%
\bibitem [{\citenamefont {Potter}\ \emph {et~al.}(2016)\citenamefont {Potter},
  \citenamefont {Morimoto},\ and\ \citenamefont {Vishwanath}}]{Potter16}%
  \BibitemOpen
  \bibfield  {author} {\bibinfo {author} {\bibfnamefont {A.~C.}\ \bibnamefont
  {Potter}}, \bibinfo {author} {\bibfnamefont {T.}~\bibnamefont {Morimoto}}, \
  and\ \bibinfo {author} {\bibfnamefont {A.}~\bibnamefont {Vishwanath}},\
  }\href {\doibase 10.1103/PhysRevX.6.041001} {\bibfield  {journal} {\bibinfo
  {journal} {Phys. Rev. X}\ }\textbf {\bibinfo {volume} {6}},\ \bibinfo {pages}
  {041001} (\bibinfo {year} {2016})}\BibitemShut {NoStop}%
\bibitem [{\citenamefont {Else}\ and\ \citenamefont {Nayak}(2016)}]{Else16}%
  \BibitemOpen
  \bibfield  {author} {\bibinfo {author} {\bibfnamefont {D.~V.}\ \bibnamefont
  {Else}}\ and\ \bibinfo {author} {\bibfnamefont {C.}~\bibnamefont {Nayak}},\
  }\href {\doibase 10.1103/PhysRevB.93.201103} {\bibfield  {journal} {\bibinfo
  {journal} {Phys. Rev. B}\ }\textbf {\bibinfo {volume} {93}},\ \bibinfo
  {pages} {201103(R)} (\bibinfo {year} {2016})}\BibitemShut {NoStop}%
\bibitem [{\citenamefont {Roy}\ and\ \citenamefont {Harper}(2016)}]{Roy16}%
  \BibitemOpen
  \bibfield  {author} {\bibinfo {author} {\bibfnamefont {R.}~\bibnamefont
  {Roy}}\ and\ \bibinfo {author} {\bibfnamefont {F.}~\bibnamefont {Harper}},\
  }\href {\doibase 10.1103/PhysRevB.94.125105} {\bibfield  {journal} {\bibinfo
  {journal} {Phys. Rev. B}\ }\textbf {\bibinfo {volume} {94}},\ \bibinfo
  {pages} {125105} (\bibinfo {year} {2016})}\BibitemShut {NoStop}%
\bibitem [{\citenamefont {von Keyserlingk}\ and\ \citenamefont
  {Sondhi}(2016)}]{Sondhi16a}%
  \BibitemOpen
  \bibfield  {author} {\bibinfo {author} {\bibfnamefont {C.~W.}\ \bibnamefont
  {von Keyserlingk}}\ and\ \bibinfo {author} {\bibfnamefont {S.~L.}\
  \bibnamefont {Sondhi}},\ }\href {\doibase 10.1103/PhysRevB.93.245145}
  {\bibfield  {journal} {\bibinfo  {journal} {Phys. Rev. B}\ }\textbf {\bibinfo
  {volume} {93}},\ \bibinfo {pages} {245145} (\bibinfo {year}
  {2016})}\BibitemShut {NoStop}%
\bibitem [{\citenamefont {Chandran}\ \emph {et~al.}(2014)\citenamefont
  {Chandran}, \citenamefont {Khemani}, \citenamefont {Laumann},\ and\
  \citenamefont {Sondhi}}]{Chandran14}%
  \BibitemOpen
  \bibfield  {author} {\bibinfo {author} {\bibfnamefont {A.}~\bibnamefont
  {Chandran}}, \bibinfo {author} {\bibfnamefont {V.}~\bibnamefont {Khemani}},
  \bibinfo {author} {\bibfnamefont {C.~R.}\ \bibnamefont {Laumann}}, \ and\
  \bibinfo {author} {\bibfnamefont {S.~L.}\ \bibnamefont {Sondhi}},\ }\href
  {\doibase 10.1103/PhysRevB.89.144201} {\bibfield  {journal} {\bibinfo
  {journal} {Phys. Rev. B}\ }\textbf {\bibinfo {volume} {89}},\ \bibinfo
  {pages} {144201} (\bibinfo {year} {2014})}\BibitemShut {NoStop}%
\bibitem [{\citenamefont {Lazarides}\ \emph {et~al.}(2014)\citenamefont
  {Lazarides}, \citenamefont {Das},\ and\ \citenamefont
  {Moessner}}]{Lazarides14}%
  \BibitemOpen
  \bibfield  {author} {\bibinfo {author} {\bibfnamefont {A.}~\bibnamefont
  {Lazarides}}, \bibinfo {author} {\bibfnamefont {A.}~\bibnamefont {Das}}, \
  and\ \bibinfo {author} {\bibfnamefont {R.}~\bibnamefont {Moessner}},\ }\href
  {\doibase 10.1103/PhysRevE.90.012110} {\bibfield  {journal} {\bibinfo
  {journal} {Phys. Rev. E}\ }\textbf {\bibinfo {volume} {90}},\ \bibinfo
  {pages} {012110} (\bibinfo {year} {2014})}\BibitemShut {NoStop}%
\bibitem [{\citenamefont {Kim}\ \emph {et~al.}(2014)\citenamefont {Kim},
  \citenamefont {Ikeda},\ and\ \citenamefont {Huse}}]{Hyungwon14}%
  \BibitemOpen
  \bibfield  {author} {\bibinfo {author} {\bibfnamefont {H.}~\bibnamefont
  {Kim}}, \bibinfo {author} {\bibfnamefont {T.~N.}\ \bibnamefont {Ikeda}}, \
  and\ \bibinfo {author} {\bibfnamefont {D.~A.}\ \bibnamefont {Huse}},\ }\href
  {\doibase 10.1103/PhysRevE.90.052105} {\bibfield  {journal} {\bibinfo
  {journal} {Phys. Rev. E}\ }\textbf {\bibinfo {volume} {90}},\ \bibinfo
  {pages} {052105} (\bibinfo {year} {2014})}\BibitemShut {NoStop}%
\bibitem [{\citenamefont {D'Alessio}\ and\ \citenamefont
  {Rigol}(2014)}]{DAlessio14}%
  \BibitemOpen
  \bibfield  {author} {\bibinfo {author} {\bibfnamefont {L.}~\bibnamefont
  {D'Alessio}}\ and\ \bibinfo {author} {\bibfnamefont {M.}~\bibnamefont
  {Rigol}},\ }\href {\doibase 10.1103/PhysRevX.4.041048} {\bibfield  {journal}
  {\bibinfo  {journal} {Phys. Rev. X}\ }\textbf {\bibinfo {volume} {4}},\
  \bibinfo {pages} {041048} (\bibinfo {year} {2014})}\BibitemShut {NoStop}%
\bibitem [{\citenamefont {Ponte}\ \emph {et~al.}(2015)\citenamefont {Ponte},
  \citenamefont {Chandran}, \citenamefont {Papić},\ and\ \citenamefont
  {Abanin}}]{Ponte15}%
  \BibitemOpen
  \bibfield  {author} {\bibinfo {author} {\bibfnamefont {P.}~\bibnamefont
  {Ponte}}, \bibinfo {author} {\bibfnamefont {A.}~\bibnamefont {Chandran}},
  \bibinfo {author} {\bibfnamefont {Z.}~\bibnamefont {Papić}}, \ and\ \bibinfo
  {author} {\bibfnamefont {D.~A.}\ \bibnamefont {Abanin}},\ }\href {\doibase
  https://doi.org/10.1016/j.aop.2014.11.008} {\bibfield  {journal} {\bibinfo
  {journal} {Annals of Physics}\ }\textbf {\bibinfo {volume} {353}},\ \bibinfo
  {pages} {196 } (\bibinfo {year} {2015})}\BibitemShut {NoStop}%
\bibitem [{\citenamefont {Haldar}\ \emph {et~al.}(2018)\citenamefont {Haldar},
  \citenamefont {Moessner},\ and\ \citenamefont {Das}}]{Haldar18}%
  \BibitemOpen
  \bibfield  {author} {\bibinfo {author} {\bibfnamefont {A.}~\bibnamefont
  {Haldar}}, \bibinfo {author} {\bibfnamefont {R.}~\bibnamefont {Moessner}}, \
  and\ \bibinfo {author} {\bibfnamefont {A.}~\bibnamefont {Das}},\ }\href
  {\doibase 10.1103/PhysRevB.97.245122} {\bibfield  {journal} {\bibinfo
  {journal} {Phys. Rev. B}\ }\textbf {\bibinfo {volume} {97}},\ \bibinfo
  {pages} {245122} (\bibinfo {year} {2018})}\BibitemShut {NoStop}%
\bibitem [{\citenamefont {Gritsev}\ and\ \citenamefont
  {Polkovnikov}(2017)}]{Gritsev17}%
  \BibitemOpen
  \bibfield  {author} {\bibinfo {author} {\bibfnamefont {V.}~\bibnamefont
  {Gritsev}}\ and\ \bibinfo {author} {\bibfnamefont {A.}~\bibnamefont
  {Polkovnikov}},\ }\href {\doibase 10.21468/SciPostPhys.2.3.021} {\bibfield
  {journal} {\bibinfo  {journal} {SciPost Phys.}\ }\textbf {\bibinfo {volume}
  {2}},\ \bibinfo {pages} {021} (\bibinfo {year} {2017})}\BibitemShut {NoStop}%
\bibitem [{\citenamefont {Asb\'oth}\ \emph {et~al.}(2014)\citenamefont
  {Asb\'oth}, \citenamefont {Tarasinski},\ and\ \citenamefont
  {Delplace}}]{Delplace14}%
  \BibitemOpen
  \bibfield  {author} {\bibinfo {author} {\bibfnamefont {J.~K.}\ \bibnamefont
  {Asb\'oth}}, \bibinfo {author} {\bibfnamefont {B.}~\bibnamefont
  {Tarasinski}}, \ and\ \bibinfo {author} {\bibfnamefont {P.}~\bibnamefont
  {Delplace}},\ }\href {\doibase 10.1103/PhysRevB.90.125143} {\bibfield
  {journal} {\bibinfo  {journal} {Phys. Rev. B}\ }\textbf {\bibinfo {volume}
  {90}},\ \bibinfo {pages} {125143} (\bibinfo {year} {2014})}\BibitemShut
  {NoStop}%
\bibitem [{\citenamefont {Yates}\ and\ \citenamefont {Mitra}(2017)}]{Yates17}%
  \BibitemOpen
  \bibfield  {author} {\bibinfo {author} {\bibfnamefont {D.~J.}\ \bibnamefont
  {Yates}}\ and\ \bibinfo {author} {\bibfnamefont {A.}~\bibnamefont {Mitra}},\
  }\href {\doibase 10.1103/PhysRevB.96.115108} {\bibfield  {journal} {\bibinfo
  {journal} {Phys. Rev. B}\ }\textbf {\bibinfo {volume} {96}},\ \bibinfo
  {pages} {115108} (\bibinfo {year} {2017})}\BibitemShut {NoStop}%
\bibitem [{\citenamefont {Yates}\ \emph {et~al.}(2018)\citenamefont {Yates},
  \citenamefont {Lemonik},\ and\ \citenamefont {Mitra}}]{Yates18}%
  \BibitemOpen
  \bibfield  {author} {\bibinfo {author} {\bibfnamefont {D.}~\bibnamefont
  {Yates}}, \bibinfo {author} {\bibfnamefont {Y.}~\bibnamefont {Lemonik}}, \
  and\ \bibinfo {author} {\bibfnamefont {A.}~\bibnamefont {Mitra}},\ }\href
  {\doibase 10.1103/PhysRevLett.121.076802} {\bibfield  {journal} {\bibinfo
  {journal} {Phys. Rev. Lett.}\ }\textbf {\bibinfo {volume} {121}},\ \bibinfo
  {pages} {076802} (\bibinfo {year} {2018})}\BibitemShut {NoStop}%
\bibitem [{\citenamefont {D'Alessio}\ and\ \citenamefont
  {Polkovnikov}(2013)}]{Polkov13}%
  \BibitemOpen
  \bibfield  {author} {\bibinfo {author} {\bibfnamefont {L.}~\bibnamefont
  {D'Alessio}}\ and\ \bibinfo {author} {\bibfnamefont {A.}~\bibnamefont
  {Polkovnikov}},\ }\href {\doibase https://doi.org/10.1016/j.aop.2013.02.011}
  {\bibfield  {journal} {\bibinfo  {journal} {Annals of Physics}\ }\textbf
  {\bibinfo {volume} {333}},\ \bibinfo {pages} {19 } (\bibinfo {year}
  {2013})}\BibitemShut {NoStop}%
\bibitem [{\citenamefont {Eckardt}\ and\ \citenamefont
  {Anisimovas}(2015)}]{Eckardt15}%
  \BibitemOpen
  \bibfield  {author} {\bibinfo {author} {\bibfnamefont {A.}~\bibnamefont
  {Eckardt}}\ and\ \bibinfo {author} {\bibfnamefont {E.}~\bibnamefont
  {Anisimovas}},\ }\href {http://stacks.iop.org/1367-2630/17/i=9/a=093039}
  {\bibfield  {journal} {\bibinfo  {journal} {New Journal of Physics}\ }\textbf
  {\bibinfo {volume} {17}},\ \bibinfo {pages} {093039} (\bibinfo {year}
  {2015})}\BibitemShut {NoStop}%
\bibitem [{\citenamefont {Abanin}\ \emph {et~al.}(2015)\citenamefont {Abanin},
  \citenamefont {De~Roeck},\ and\ \citenamefont {Huveneers}}]{Abanin15}%
  \BibitemOpen
  \bibfield  {author} {\bibinfo {author} {\bibfnamefont {D.~A.}\ \bibnamefont
  {Abanin}}, \bibinfo {author} {\bibfnamefont {W.}~\bibnamefont {De~Roeck}}, \
  and\ \bibinfo {author} {\bibfnamefont {F.}\ \bibnamefont {Huveneers}},\
  }\href {\doibase 10.1103/PhysRevLett.115.256803} {\bibfield  {journal}
  {\bibinfo  {journal} {Phys. Rev. Lett.}\ }\textbf {\bibinfo {volume} {115}},\
  \bibinfo {pages} {256803} (\bibinfo {year} {2015})}\BibitemShut {NoStop}%
\bibitem [{\citenamefont {Bertini}\ \emph {et~al.}(2015)\citenamefont
  {Bertini}, \citenamefont {Essler}, \citenamefont {Groha},\ and\ \citenamefont
  {Robinson}}]{Bertini15}%
  \BibitemOpen
  \bibfield  {author} {\bibinfo {author} {\bibfnamefont {B.}~\bibnamefont
  {Bertini}}, \bibinfo {author} {\bibfnamefont {F.~H.~L.}\ \bibnamefont
  {Essler}}, \bibinfo {author} {\bibfnamefont {S.}~\bibnamefont {Groha}}, \
  and\ \bibinfo {author} {\bibfnamefont {N.~J.}\ \bibnamefont {Robinson}},\
  }\href {\doibase 10.1103/PhysRevLett.115.180601} {\bibfield  {journal}
  {\bibinfo  {journal} {Phys. Rev. Lett.}\ }\textbf {\bibinfo {volume} {115}},\
  \bibinfo {pages} {180601} (\bibinfo {year} {2015})}\BibitemShut {NoStop}%
\bibitem [{\citenamefont {Mitra}(2018)}]{AMreview}%
  \BibitemOpen
  \bibfield  {author} {\bibinfo {author} {\bibfnamefont {A.}~\bibnamefont
  {Mitra}},\ }\href@noop {} {\bibfield  {journal} {\bibinfo  {journal} {Annual
  Review of Condensed Matter Physics}\ }\textbf {\bibinfo {volume} {9}},\
  \bibinfo {pages} {245} (\bibinfo {year} {2018})}\BibitemShut {NoStop}%
\bibitem [{\citenamefont {Kuwahara}\ \emph {et~al.}(2016)\citenamefont
  {Kuwahara}, \citenamefont {Mori},\ and\ \citenamefont
  {Saito}}]{Kuwahara2016}%
  \BibitemOpen
  \bibfield  {author} {\bibinfo {author} {\bibfnamefont {T.}~\bibnamefont
  {Kuwahara}}, \bibinfo {author} {\bibfnamefont {T.}~\bibnamefont {Mori}}, \
  and\ \bibinfo {author} {\bibfnamefont {K.}~\bibnamefont {Saito}},\ }\href
  {\doibase https://doi.org/10.1016/j.aop.2016.01.012} {\bibfield  {journal}
  {\bibinfo  {journal} {Annals of Physics}\ }\textbf {\bibinfo {volume}
  {367}},\ \bibinfo {pages} {96 } (\bibinfo {year} {2016})}\BibitemShut
  {NoStop}%
\bibitem [{\citenamefont {Bukov}\ \emph {et~al.}(2016)\citenamefont {Bukov},
  \citenamefont {Heyl}, \citenamefont {Huse},\ and\ \citenamefont
  {Polkovnikov}}]{Bukov16}%
  \BibitemOpen
  \bibfield  {author} {\bibinfo {author} {\bibfnamefont {M.}~\bibnamefont
  {Bukov}}, \bibinfo {author} {\bibfnamefont {M.}~\bibnamefont {Heyl}},
  \bibinfo {author} {\bibfnamefont {D.~A.}\ \bibnamefont {Huse}}, \ and\
  \bibinfo {author} {\bibfnamefont {A.}~\bibnamefont {Polkovnikov}},\
  }\href@noop {} {\bibfield  {journal} {\bibinfo  {journal} {Phys. Rev. B}\
  }\textbf {\bibinfo {volume} {93}},\ \bibinfo {pages} {155132} (\bibinfo
  {year} {2016})}\BibitemShut {NoStop}%
\bibitem [{\citenamefont {Mori}\ \emph {et~al.}(2016)\citenamefont {Mori},
  \citenamefont {Kuwahara},\ and\ \citenamefont {Saito}}]{Mori16}%
  \BibitemOpen
  \bibfield  {author} {\bibinfo {author} {\bibfnamefont {T.}~\bibnamefont
  {Mori}}, \bibinfo {author} {\bibfnamefont {T.}~\bibnamefont {Kuwahara}}, \
  and\ \bibinfo {author} {\bibfnamefont {K.}~\bibnamefont {Saito}},\ }\href
  {\doibase 10.1103/PhysRevLett.116.120401} {\bibfield  {journal} {\bibinfo
  {journal} {Phys. Rev. Lett.}\ }\textbf {\bibinfo {volume} {116}},\ \bibinfo
  {pages} {120401} (\bibinfo {year} {2016})}\BibitemShut {NoStop}%
\bibitem [{\citenamefont {Else}\ \emph
  {et~al.}(2017{\natexlab{b}})\citenamefont {Else}, \citenamefont {Bauer},\
  and\ \citenamefont {Nayak}}]{Else17}%
  \BibitemOpen
  \bibfield  {author} {\bibinfo {author} {\bibfnamefont {D.~V.}\ \bibnamefont
  {Else}}, \bibinfo {author} {\bibfnamefont {B.}~\bibnamefont {Bauer}}, \ and\
  \bibinfo {author} {\bibfnamefont {C.}~\bibnamefont {Nayak}},\ }\href
  {\doibase 10.1103/PhysRevX.7.011026} {\bibfield  {journal} {\bibinfo
  {journal} {Phys. Rev. X}\ }\textbf {\bibinfo {volume} {7}},\ \bibinfo {pages}
  {011026} (\bibinfo {year} {2017}{\natexlab{b}})}\BibitemShut {NoStop}%
\bibitem [{\citenamefont {Abanin}\ \emph
  {et~al.}(2017{\natexlab{a}})\citenamefont {Abanin}, \citenamefont {De~Roeck},
  \citenamefont {Ho},\ and\ \citenamefont {Huveneers}}]{Abanin17}%
  \BibitemOpen
  \bibfield  {author} {\bibinfo {author} {\bibfnamefont {D.~A.}\ \bibnamefont
  {Abanin}}, \bibinfo {author} {\bibfnamefont {W.}~\bibnamefont {De~Roeck}},
  \bibinfo {author} {\bibfnamefont {W.~W.}\ \bibnamefont {Ho}}, \ and\ \bibinfo
  {author} {\bibfnamefont {F.}\ \bibnamefont {Huveneers}},\ }\href
  {\doibase 10.1103/PhysRevB.95.014112} {\bibfield  {journal} {\bibinfo
  {journal} {Phys. Rev. B}\ }\textbf {\bibinfo {volume} {95}},\ \bibinfo
  {pages} {014112} (\bibinfo {year} {2017}{\natexlab{a}})}\BibitemShut
  {NoStop}%
\bibitem [{\citenamefont {Abanin}\ \emph
  {et~al.}(2017{\natexlab{b}})\citenamefont {Abanin}, \citenamefont {De~Roeck},
  \citenamefont {Ho},\ and\ \citenamefont {Huveneers}}]{Abanin17b}%
  \BibitemOpen
  \bibfield  {author} {\bibinfo {author} {\bibfnamefont {D.}~\bibnamefont
  {Abanin}}, \bibinfo {author} {\bibfnamefont {W.}~\bibnamefont {De~Roeck}},
  \bibinfo {author} {\bibfnamefont {W.~W.}\ \bibnamefont {Ho}}, \ and\ \bibinfo
  {author} {\bibfnamefont {F.}~\bibnamefont {Huveneers}},\ }\href {\doibase
  10.1007/s00220-017-2930-x} {\bibfield  {journal} {\bibinfo  {journal}
  {Communications in Mathematical Physics}\ }\textbf {\bibinfo {volume}
  {354}},\ \bibinfo {pages} {809} (\bibinfo {year}
  {2017}{\natexlab{b}})}\BibitemShut {NoStop}%
\bibitem [{\citenamefont {Essler}\ and\ \citenamefont
  {Fagotti}(2016)}]{Essler16}%
  \BibitemOpen
  \bibfield  {author} {\bibinfo {author} {\bibfnamefont {F.~H.~L.}\
  \bibnamefont {Essler}}\ and\ \bibinfo {author} {\bibfnamefont
  {M.}~\bibnamefont {Fagotti}},\ }\href {\doibase
  10.1088/1742-5468/2016/06/064002} {\bibfield  {journal} {\bibinfo  {journal}
  {Journal of Statistical Mechanics: Theory and Experiment}\ }\textbf {\bibinfo
  {volume} {2016}},\ \bibinfo {pages} {064002} (\bibinfo {year}
  {2016})}\BibitemShut {NoStop}%
\bibitem [{\citenamefont {Vajna}\ \emph {et~al.}(2018)\citenamefont {Vajna},
  \citenamefont {Klobas}, \citenamefont {Prosen},\ and\ \citenamefont
  {Polkovnikov}}]{Prosen18}%
  \BibitemOpen
  \bibfield  {author} {\bibinfo {author} {\bibfnamefont {S.}~\bibnamefont
  {Vajna}}, \bibinfo {author} {\bibfnamefont {K.}~\bibnamefont {Klobas}},
  \bibinfo {author} {\bibfnamefont {T.}\ \bibnamefont {Prosen}}, \ and\
  \bibinfo {author} {\bibfnamefont {A.}~\bibnamefont {Polkovnikov}},\ }\href
  {\doibase 10.1103/PhysRevLett.120.200607} {\bibfield  {journal} {\bibinfo
  {journal} {Phys. Rev. Lett.}\ }\textbf {\bibinfo {volume} {120}},\ \bibinfo
  {pages} {200607} (\bibinfo {year} {2018})}\BibitemShut {NoStop}%
\bibitem [{\citenamefont {Jiang}\ \emph {et~al.}(2011)\citenamefont {Jiang},
  \citenamefont {Kitagawa}, \citenamefont {Alicea}, \citenamefont {Akhmerov},
  \citenamefont {Pekker}, \citenamefont {Refael}, \citenamefont {Cirac},
  \citenamefont {Demler}, \citenamefont {Lukin},\ and\ \citenamefont
  {Zoller}}]{Zoller11}%
  \BibitemOpen
  \bibfield  {author} {\bibinfo {author} {\bibfnamefont {L.}~\bibnamefont
  {Jiang}}, \bibinfo {author} {\bibfnamefont {T.}~\bibnamefont {Kitagawa}},
  \bibinfo {author} {\bibfnamefont {J.}~\bibnamefont {Alicea}}, \bibinfo
  {author} {\bibfnamefont {A.~R.}\ \bibnamefont {Akhmerov}}, \bibinfo {author}
  {\bibfnamefont {D.}~\bibnamefont {Pekker}}, \bibinfo {author} {\bibfnamefont
  {G.}~\bibnamefont {Refael}}, \bibinfo {author} {\bibfnamefont {J.~I.}\
  \bibnamefont {Cirac}}, \bibinfo {author} {\bibfnamefont {E.}~\bibnamefont
  {Demler}}, \bibinfo {author} {\bibfnamefont {M.~D.}\ \bibnamefont {Lukin}}, \
  and\ \bibinfo {author} {\bibfnamefont {P.}~\bibnamefont {Zoller}},\ }\href
  {\doibase 10.1103/PhysRevLett.106.220402} {\bibfield  {journal} {\bibinfo
  {journal} {Phys. Rev. Lett.}\ }\textbf {\bibinfo {volume} {106}},\ \bibinfo
  {pages} {220402} (\bibinfo {year} {2011})}\BibitemShut {NoStop}%
\bibitem [{\citenamefont {Benito}\ \emph {et~al.}(2014)\citenamefont {Benito},
  \citenamefont {G\'omez-Le\'on}, \citenamefont {Bastidas}, \citenamefont
  {Brandes},\ and\ \citenamefont {Platero}}]{Benito14}%
  \BibitemOpen
  \bibfield  {author} {\bibinfo {author} {\bibfnamefont {M.}~\bibnamefont
  {Benito}}, \bibinfo {author} {\bibfnamefont {A.}~\bibnamefont
  {G\'omez-Le\'on}}, \bibinfo {author} {\bibfnamefont {V.~M.}\ \bibnamefont
  {Bastidas}}, \bibinfo {author} {\bibfnamefont {T.}~\bibnamefont {Brandes}}, \
  and\ \bibinfo {author} {\bibfnamefont {G.}~\bibnamefont {Platero}},\ }\href
  {\doibase 10.1103/PhysRevB.90.205127} {\bibfield  {journal} {\bibinfo
  {journal} {Phys. Rev. B}\ }\textbf {\bibinfo {volume} {90}},\ \bibinfo
  {pages} {205127} (\bibinfo {year} {2014})}\BibitemShut {NoStop}%
\bibitem [{\citenamefont {Berdanier}\ \emph {et~al.}(2017)\citenamefont
  {Berdanier}, \citenamefont {Kolodrubetz}, \citenamefont {Vasseur},\ and\
  \citenamefont {Moore}}]{Berdanier17}%
  \BibitemOpen
  \bibfield  {author} {\bibinfo {author} {\bibfnamefont {W.}~\bibnamefont
  {Berdanier}}, \bibinfo {author} {\bibfnamefont {M.}~\bibnamefont
  {Kolodrubetz}}, \bibinfo {author} {\bibfnamefont {R.}~\bibnamefont
  {Vasseur}}, \ and\ \bibinfo {author} {\bibfnamefont {J.~E.}\ \bibnamefont
  {Moore}},\ }\href {\doibase 10.1103/PhysRevLett.118.260602} {\bibfield
  {journal} {\bibinfo  {journal} {Phys. Rev. Lett.}\ }\textbf {\bibinfo
  {volume} {118}},\ \bibinfo {pages} {260602} (\bibinfo {year}
  {2017})}\BibitemShut {NoStop}%
\bibitem [{\citenamefont {Berdanier}\ \emph {et~al.}(2018)\citenamefont
  {Berdanier}, \citenamefont {Kolodrubetz}, \citenamefont {Parameswaran},\ and\
  \citenamefont {Vasseur}}]{Berdanier18}%
  \BibitemOpen
  \bibfield  {author} {\bibinfo {author} {\bibfnamefont {W.}~\bibnamefont
  {Berdanier}}, \bibinfo {author} {\bibfnamefont {M.}~\bibnamefont
  {Kolodrubetz}}, \bibinfo {author} {\bibfnamefont {S.~A.}\ \bibnamefont
  {Parameswaran}}, \ and\ \bibinfo {author} {\bibfnamefont {R.}~\bibnamefont
  {Vasseur}},\ }\href {\doibase 10.1073/pnas.1805796115} {\bibfield  {journal}
  {\bibinfo  {journal} {Proceedings of the National Academy of Sciences}\
  }\textbf {\bibinfo {volume} {115}},\ \bibinfo {pages} {9491} (\bibinfo {year}
  {2018})}\BibitemShut {NoStop}%
\bibitem [{\citenamefont {Fidkowski}\ and\ \citenamefont
  {Kitaev}(2011)}]{Fidk11}%
  \BibitemOpen
  \bibfield  {author} {\bibinfo {author} {\bibfnamefont {L.}~\bibnamefont
  {Fidkowski}}\ and\ \bibinfo {author} {\bibfnamefont {A.}~\bibnamefont
  {Kitaev}},\ }\href {\doibase 10.1103/PhysRevB.83.075103} {\bibfield
  {journal} {\bibinfo  {journal} {Phys. Rev. B}\ }\textbf {\bibinfo {volume}
  {83}},\ \bibinfo {pages} {075103} (\bibinfo {year} {2011})}\BibitemShut
  {NoStop}%
\bibitem [{\citenamefont {Verresen}\ \emph {et~al.}(2017)\citenamefont
  {Verresen}, \citenamefont {Moessner},\ and\ \citenamefont
  {Pollmann}}]{Verresen17}%
  \BibitemOpen
  \bibfield  {author} {\bibinfo {author} {\bibfnamefont {R.}~\bibnamefont
  {Verresen}}, \bibinfo {author} {\bibfnamefont {R.}~\bibnamefont {Moessner}},
  \ and\ \bibinfo {author} {\bibfnamefont {F.}~\bibnamefont {Pollmann}},\
  }\href {\doibase 10.1103/PhysRevB.96.165124} {\bibfield  {journal} {\bibinfo
  {journal} {Phys. Rev. B}\ }\textbf {\bibinfo {volume} {96}},\ \bibinfo
  {pages} {165124} (\bibinfo {year} {2017})}\BibitemShut {NoStop}%
\end{thebibliography}


%

\end{document}